\documentclass[conference]{IEEEtran}
\ifCLASSINFOpdf
  \usepackage[pdftex]{graphicx}
  \graphicspath{{../figures/}}
  \DeclareGraphicsExtensions{.pdf,.jpeg,.png}
\else
\fi
\usepackage[hyphens]{url}

\usepackage[dvipsnames]{xcolor} 
\usepackage{booktabs} 
\usepackage{tikz} 
\usepackage{pgfplots} 
\usepackage{multicol} 

\usepackage[utf8]{inputenc}
\usepackage[T1]{fontenc}
\usepackage[bulgarian,greek,english]{babel}
\usepackage{csquotes} 
\usepackage{subfig}
\usepackage{balance}
\usepackage{dblfloatfix}
\usepackage{threeparttable}
\usepackage{tabularx}
\usepackage{hyperref}
\usepackage{breakurl}

\newcommand{\circfill}[1][black,fill=black]{\tikz[baseline=-0.5ex]\draw[#1,radius=2.5pt] (0,0) circle ;}

\newcommand{\circempty}[1][black,fill=none]{\tikz[baseline=-0.5ex]\draw[#1,radius=2.5pt] (0,0) circle ;}

\newcommand*\rot{\rotatebox{90}}

\newcommand{\ie}{i.\,e.}
\newcommand{\eg}{e.\,g.}

\newcommand{\numCBlibs}{28\ }

\newcommand\fixme[1]{{{\color{black}#1}}}

\usepackage{soul}
\newcommand{\claim}[1]{\hl{#1}}

\renewcommand{\hl}[1]{{\color{black}#1}}

\begin{document}
%
\title{\textit{We Value Your Privacy ... Now Take Some Cookies:} Measuring the GDPR’s Impact on Web Privacy}



%
\author{\IEEEauthorblockN{Martin Degeling\IEEEauthorrefmark{1},
Christine Utz\IEEEauthorrefmark{1},
Christopher Lentzsch\IEEEauthorrefmark{1}, 
Henry Hosseini\IEEEauthorrefmark{1},
Florian Schaub\IEEEauthorrefmark{2}, and
Thorsten Holz\IEEEauthorrefmark{1}}
\IEEEauthorblockA{\IEEEauthorrefmark{1}Ruhr-Universität Bochum, Germany\\
Email: \{firstname.lastname\}@rub.de}
\IEEEauthorblockA{\IEEEauthorrefmark{2}University of Michigan, Ann Arbor, MI, USA\\
Email: fschaub@umich.edu}}


\IEEEoverridecommandlockouts
\makeatletter\def\@IEEEpubidpullup{6.5\baselineskip}\makeatother
\IEEEpubid{\parbox{\columnwidth}{
    Network and Distributed Systems Security (NDSS) Symposium 2019\\
    24-27 February 2019, San Diego, CA, USA\\
    ISBN 1-891562-55-X\\
    https://dx.doi.org/10.14722/ndss.2019.23xxx\\
    www.ndss-symposium.org
}
\hspace{\columnsep}\makebox[\columnwidth]{}}

\maketitle

%
%

\begin{abstract}
The European Union's General Data Protection Regulation (GDPR) went into effect on May 25, 2018. Its privacy regulations apply to any service and company collecting or processing personal data in Europe. Many companies had to adjust their data handling processes, consent forms, and privacy policies to comply with the GDPR's transparency requirements. We monitored this rare event 
\claim{by analyzing changes}  
on popular websites in all 28 member states of the European Union. For each country, we periodically examined its 500 most popular websites -- 6,579 in total -- for the presence of and updates to their privacy policy \hl{between December 2017 and October 2018}. 
%
While many websites already had privacy policies, 
we find that in some countries up to 15.7\,\% of websites added new privacy policies by May 25, 2018, resulting in 84.5\,\% of websites having privacy policies. 72.6\,\% of websites with existing privacy policies updated them close to the date. \hl{After May this positive development slowed down noticeably.}
Most visibly, 62.1\,\% of websites in Europe now display cookie consent notices, 16\,\% more than in January 2018. These 
notices inform users about a site's cookie use and user tracking practices. We categorized all observed 
cookie consent notices and evaluated \numCBlibs common implementations with respect to their technical realization of cookie consent. Our analysis shows that core web security mechanisms such as the same-origin policy pose problems for the implementation of consent according to GDPR rules, and opting out of third-party cookies requires the third party to cooperate. Overall, we conclude that \fixme{the web became more transparent at the time GDPR came into force}, but there is still a lack of both functional and usable mechanisms for users to consent to or deny processing of their personal data on the Internet. 
\end{abstract}

\section{Introduction}
On May 25, 2018, the General Data Protection Regulation (GDPR) went into effect in the European Union. The GDPR is supposed to set high and consistent standards for the processing of personal data within the European Union and whenever personal data of people residing in Europe is involved. 
As a result, the GDPR affects millions of web services from around the world which are available in Europe. In addition to potentially changing how they process personal data, companies have to disclose transparently how they handle personal data, the legal bases for their data processing, and need to offer their users mechanisms for individual consent, data access, data deletion, and data portability. Even outside Europe, online services had to prepare for the GDPR because it 
not only applies to companies in Europe but any company that offers its service in Europe. As a result, the GDPR is expected to have a major impact on companies across the world.

Previous work has found that about 70 to 80\,\% of websites in the U.S. have privacy policies~\cite{liu_raising_2002,nokhbeh_zaeem_study_2017}. But analysis of privacy policies has been focused on \hl{English-language} policies, performing in-depth studies on their content~\cite{wilson_creation_2016,harkous_polisis:_2018,libert_2018,tesfay_privacyguide_2018}. Cookie consent notices have just recently seen research attention with respect to their usability~\cite{oksana_kulyk_this_2018}, but their use and implementations have not been studied in detail, yet.

In this paper, we describe an empirical study to measure \fixme{changes that occurred on a representative set of websites at the time the GDPR came into force}. We monitored this rare event by analyzing the 500 most visited websites, according to Alexa country rankings, in each of the 28 member states of the EU over the course of \hl{eleven} months. In total, this resulted in a set of 6,759 websites available in 24 different languages.
We used a combination of automated and manual methods and compared the privacy policies of these websites before and after the GDPR enforcement date and, together with historic data, retrieved 112,041 privacy policies. 

Our results show that \fixme{changes made around the GDPR enforcement date} had overall positive effect on the transparency of websites: more websites (+4.9\,\%) now have privacy policies and/or inform users about their cookie practices and increasingly inform users about their rights and the legal basis of their data processing. But even though on average  84.5\,\% of the websites we checked for each country now have privacy policies, differences remain high. 
By tracing the changes on policies, we found that, despite the GDPR's two-year grace period, 50\,\% of websites updated their privacy policies in May 2018 just before the GDPR went into effect, and more than 60\,\% did not make any change in 2016 or 2017.
We further found that actual practices did not change much: The amount of tracking stayed the same and the majority of sites relies on opt-out consent mechanisms. We identified only 37 sites that asked for explicit consent before setting cookies.

For web users in Europe, the most visible change is an increase in cookie consent notices and the features they offer, \eg, 
specific user choices for tracking and social media cookies. On average, 62.1\,\% of the analyzed websites now use such cookie banners (46.1\,\% in January 2018).
In order to better understand this phenomenon, we manually inspected 9,044 domains for their use of cookie banners and evaluated \numCBlibs common cookie consent libraries for features useful for the implementation of GDPR-compliant consent.
We found that existing implementations greatly vary in functionality, especially the granularity of control offered to the user and the ability to apply the desired cookie configuration.

In summary, our paper makes the following contributions:
\begin{enumerate}
\item We conduct an empirical, longitudinal study of privacy policies and cookie consent notices of 6,759 websites representing the 500 most popular websites in each of the 28 member states of the EU. From January to \hl{October} 2018, 
we performed monthly scans to measure changes in adoption rates. Between January and the end of May, we observed an average rise of websites providing privacy policies by 4,9 percentage points and cookie consent notices by 16. \hl{After May the development slowed down: Between June and November, the number of websites that added privacy policies and cookie consent notices increased by 0.9 and 1.1 percentage points, respectively.}
\item 
While prior studies primarily focused on English-language privacy policies, we analyze privacy policies in 24 different languages. We use natural language processing techniques to identify how privacy policies' content has changed and whether the GDPR's new transparency requirements are reflected in the texts. We find that not too many websites make use of GDPR terminology, but for those that do, the amount of information about users' rights and the legal basis of processing increased.
\item We compare the use of cookies and third-party libraries in our set of websites between January and June 2018 to determine whether the GDPR's transparency and consent requirements affected the prevalence of web tracking. While both were not significantly impacted, 147 sites stopped using tracking libraries and 37 chose to ask for explicit consent before activating them.
\item We categorize observed cookie consent notices based on their options for interaction. In our data set, we found many distinct implementations of cookie consent notices. We analyze these libraries for key features required to implement the GDPR notion of ``informed consent'' and identify technical obstacles to achieving this goal.
\end{enumerate}

\begingroup
\let\clearpage\relax
\section{Background}

\label{sec:background}
As background, we discuss the GDPR's legal requirements and technical aspects of their implementation.
\subsection{Legal Background}

In 2012, the EU started to take regulatory action to harmonize data protection laws across its member states. Existing data protection legislation comprised the Data Protection Directive (95/46/EC) \cite{data_protection_directive_1995} and the ePrivacy Directive (2002/58/EC) \cite{eprivacy_directive_2002}, along with national laws in the EU member countries implementing the requirements of the two directives.\footnote{In contrast to EU regulations, which are directly applicable in each member state, EU directives are only binding as to the result leaving the member states to decide upon the form and methods for achieving the aim.}

As pointed out by Recital 9 of the GDPR, these national implementations differed widely, resulting in a complex landscape of privacy laws across Europe. Some member states embraced stricter privacy laws and enforcement while others opted for lighter regulation.
The General Data Protection Regulation (GDPR)
\cite{gdpr_2016}
is intended to overcome this situation and harmonize privacy laws throughout the EU. It was proposed in January 2012, adopted on May 24, 2016, and its provisions became enforceable on May 25, 2018. A second regulation, the ePrivacy Regulation, is meant to complement the GDPR and complete the harmonization process. It is currently passing through the EU's legislative process.

The GDPR has several implications for web services and is therefore expected to impact the technical design of websites, what data they collect, and how they inform users about their practices. GDPR thus governs any processing of personal data for services offered in the EU, even if the service provider does not have any legal representation there. Article 3 states that the regulation applies to \textit{``the processing of personal data in the context of the activities of an establishment of a controller or a processor in the [European] Union, regardless of whether the processing takes place in the [European] Union or not.''} For online services this means that any website offering its service in the EU has to comply with GDPR standards.

Following are selected key requirements of the GDPR relevant for our study. A more detailed discussion of the
regulation can be found in legal literature \cite{ruecker_kugler_gdpr_2018}.

\textbf{Transparency.} Article 12 GDPR requires that anyone who processes personal data should inform the data subject about the fact (\eg, in a privacy policy) and present the information in \emph{``a concise, transparent, intelligible, and easily accessible form, using clear and plain language''}. Since IP addresses are considered personal data in the EU, this means that every website and the underlying web server that processes these addresses is required to provide this information.
    Article 13 more specifically lists what information needs to be provided. This includes contact data, the purposes and legal basis for the processing, and the data subject's rights regarding their personal data, \eg, the right to access, rectification, or deletion. These requirements make it necessary for every website to have a privacy policy and modify existing privacy policies to comply with the new transparency requirements.

    \textbf{Data protection by design and by default.} Article 25 states that entities processing personal data should \emph{``implement appropriate technical and organisational measures [...] designed to implement data-protection principles [...] in an effective manner''}, \emph{``taking into account [...] the state of the art''}. They are required to \emph{``ensure that by default personal data are not made accessible without the individual's intervention to an indefinite number of natural persons''}.

    Higher protection standards are required for sensitive categories of personal information like health data (Article 9).

\textbf{Consent.} According to Article 6, the processing of personal data is only lawful if one of six scenarios applies.

They include the case when the processing is necessary \emph{``for the purposes of the legitimate interests [of] the controller or [...] a third party''} (Article 6(1)(f)) or to comply with a legal obligation (Article 6(1)(c)).

Most importantly, the processing of personal data is lawful if \emph{``the data subject has given consent''} (Article 6(1)(a)). Consent, in turn, is defined in Article 2(11) as \emph{``any freely given, specific, informed and unambiguous indication of the data subject's wishes [...]''}.

Here, \emph{``freely given''} means the data subject has to be offered real choice and control; if they feel compelled to agree to the processing of their personal data, this does not constitute valid consent \cite{article_29_consent}.
For children under the age of 16 consent can only be given by the holder of parental responsibility (Article 8).

\textbf{Consent to the use of cookies.} In an earlier harmonization effort, Directive 2009/136/EC had

changed Article 5(3) of the ePrivacy Directive (2002/58/EC) to state that \emph{``the storing of information [...] in the terminal equipment of a [...] user''} is only allowed if the user \emph{``has given his or her consent, having been provided with [...] information [...] about the purposes of the processing''}~\cite{cookie_directive_2009}. This consent requirement does not apply if storing or accessing the information is \emph{``strictly necessary''} for the delivery of the service requested by the user. For websites, this is understood to exempt cookies from consent if the site would not work without setting the cookie. Examples include cookies remembering the state of the shopping cart in an online shop or the fact that the user has logged in.

This piece of legislation has caused websites across the EU to display \emph{cookie consent notices}, often referred to as \emph{cookie banners} -- boxes or banners informing users about the use of cookies by the website and associated third parties. These notices may explicitly ask users for their consent or interpret a user's continued website use as implied consent.

\hl{However, according to EU guidelines,} valid consent needs to be a freely given, active choice based on specific information about the purpose of the processing and given before the processing starts \cite{article_29_cookie_consent}.
It has to be noted that Article 5(3) applies to any kind of information stored on the user's system even if it does not contain any personal information. In case it does, consent according to GDPR rules is also required, though the two types may be merged in practice \cite{ruecker_kugler_gdpr_2018}.

\subsection{Technical Background}

\begin{figure*}
	\centering
	\includegraphics[width=1.0\textwidth]{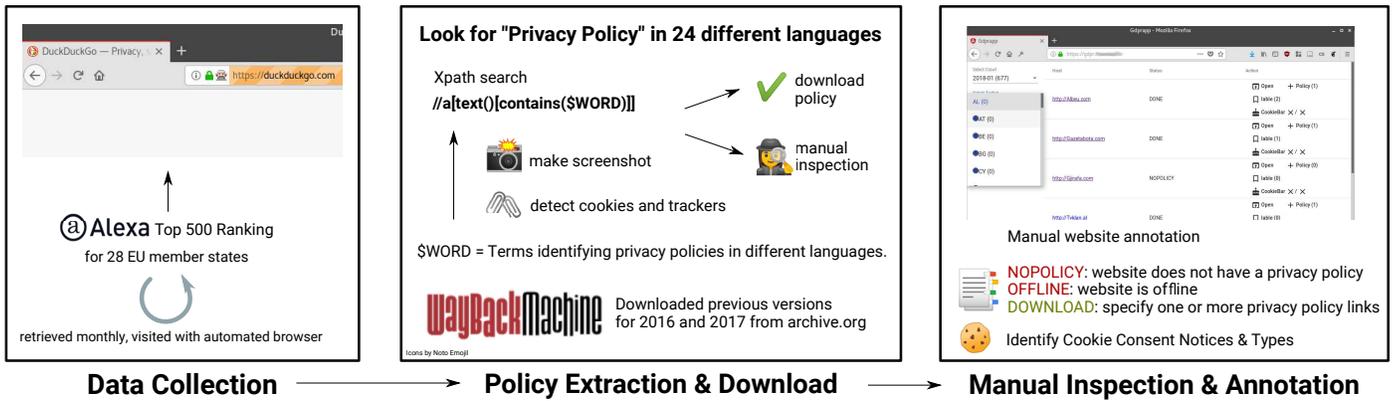}
	\caption{\hl{Overview of the website analysis process combining automated analysis, manual validation, and annotation.}}
	\label{fig:methodology}
	\vspace{-0.7cm}
\end{figure*}

Different technical solutions have been proposed to help users cope with the ever-growing number of online tracking and profiling services. In 2002, the Platform for Privacy Preferences (P3P) Project~\cite{cranor_platform_2002} was officially recommended by the W3C. It relied on machine-readable privacy policies directly interpreted by the browser,

which was enabled to automatically negotiate, \eg, the handling of certain cookies based on the user's preferences. However, none of the major web browsers support P3P anymore due to a lack of adoption by websites~\cite{cranor_necessary_2012}. Another approach is the Do Not Track (DNT) Header for the HTTP protocol, proposed in 2009~\cite{singer_tracking_17}. DNT is supported by all major browsers and allows the user to signal online content providers their preference towards tracking and behavioral advertising. However, many websites do not honor DNT signals~\cite{englehardt_cookies_2015}.

Companies in the online behavioral advertising (OBA) business point to their self-regulation program \emph{AdChoices}. The user is informed by a little blue icon in the advert and given additional information on click. The \emph{WebChoice} tool allows users to opt-out of OBA for each participating company. For users this remains challenging as studies have shown that users can hardly distinguish between different OBA companies~\cite{leon2012johnny} and have problems to even recognize and locate the corresponding icons~\cite{garlach2018adchoices}.

Apart from these solutions based on
browser settings, natural language privacy policies remain the main means to inform the user about websites' data processing practices. Studies have shown that users rarely read privacy policies because of their length and complex vocabulary~\cite{mcdonald_cost_2008, obar_biggest_2018}.

Advances in natural language processing~\cite{harkous_polisis:_2018, tesfay_privacyguide_2018} have led to the development of automated solutions to read and understand key contents of privacy policies and display them to users in an accessible fashion. However, existing solutions rely on the presence of an English-language privacy policy.

\section{Studying Privacy Policies}

To analyze the impact of GDPR enforcement on websites in the EU, we used automated tools combined with manual verification and annotation of websites in 24 different languages. We built a system to automatically scan websites for links to privacy policies, manually reviewed sites where a policy could not be extracted automatically and annotated the whole set of websites for their topic and the use of cookie consent notices. Figure \ref{fig:methodology} provides an overview of the main components of our privacy policy detection and analysis system. We describe the data collection and policy analysis method in this section, followed by the policy analysis results in Section~\ref{sec:evaluation}. Sections~\ref{sec:cbtypes} and \ref{sec:cookieeval} describe the cookie consent notice analysis and its findings.

We started by reviewing the 500 most popular websites in each of the 28 EU member states as listed by the ranking service Alexa.\footnote{\url{https://www.alexa.com/topsites}}
To extend the scope of our study, we retrieved updated top lists once per month.
\hl{After a pretest in December 2017, the websites were scanned once per month from January to April 2018, three times in May (two times before and one time after May 25, 2018) and again once per month until October 2018, resulting in 12 scans in total.}

\subsection{Automated Search for Privacy Policies}
Our automated web browser was set up in a German data center with the Selenium web driver using the latest version of Firefox (version 57 onward) on servers running Ubuntu Linux and an Xserver so that all pages were actually rendered. The results were stored in a MongoDB database. The following steps were performed for each website on its homepage after it had been completely rendered by the browser.

\textbf{Find privacy policy:}
We identified phrases pointing to privacy policies, using dictionaries and verifying the results in a prestudy. The list, \hl{which is available in our Github repository}\footnote{\url{https://github.com/RUB-SysSec/we-value-your-privacy}.}, contained phrases from all 24 official languages, plus 4 other languages spoken in the EU.

In our automated search, we only used phrases specific to privacy policies to avoid false positive results. Using an XPath query, we searched for hyperlinks that contained these phrases and saved the corresponding pages in a text file.

\textbf{Analyze website:}
We searched for domain names of third-party advertising and tracking libraries in the fully rendered page based on EasyList\footnote{See \url{https://easylist.to/easylist/easylist.txt}.}, which is often used in popular ad-blocking browser extensions.
A screenshot of the rendered homepage was made to allow for manual inspection for cookie consent notices.

Due to the complexity of websites and an often poor implementation of standards, as well as different ways of displaying long online texts such as privacy policies,
we considered a fully automated approach not sufficient to conclusively determine whether a website has a privacy policy. The word list worked well on business and news websites, but it missed privacy policy links on other sites.

Problems occurred, for example, in countries where multiple languages are spoken (\eg, Belgium, which has multiple official languages, or Estonia with its large Russian-speaking minority) as websites often present a screen asking the user to choose a language before proceeding to the actual site with its privacy policy links. Other websites did not use common phrases or would

incorporate the privacy policy into their ``terms of service''. Our system marked the websites on which automatic detection failed for manual review.
We complemented the automated search with manual validation.

\subsection{Manual Review}
\label{sec:manual_annotation}

In order to validate the results of the automated detection of privacy policies, we implemented a web-based annotation tool to review and further process the collected data. The automatic tool assigned each website one of the following status codes:

\begin{itemize}
\item \textit{Done:} A link to a privacy policy has been found and the corresponding document was downloaded (see Section \ref{sec:evaluation} for how we evaluated the content of these documents).

\item \textit{Review:} The automated analysis found word(s) from the list suggesting that a privacy policy might exist, but the system failed to download any pages.
\item \textit{No Link Found:} None of the words form the list of privacy policy identifiers was found.
\end{itemize}

All websites categorized as \textit{Review} or \textit{No Link Found} were manually inspected and annotated by the authors.

Manual inspection was done with off-the-shelf browsers and, if necessary, using Google Translate when inspecting pages in languages the annotator was unfamiliar with. Translations through Google were available in all encountered languages and good enough to figure out the general topic of a website and whether it had a privacy policy, together with

common design principles like using footers for notices and information. If a privacy policy or similar page was identified,
the policy link was added to the database, and the policy was subsequently downloaded.

If the annotator was not able to identify a privacy policy on the website, even after trying to create an account on the website, it was annotated as \textit{No Policy}. Websites that could not be reached were labeled \textit{Offline}. Under this label we merged all sites that were not reachable, occupied by a domain grabbing service, produced a screen indicating that the website was not available because of the detected location of our IP address, or belonged to a discontinued or not publicly accessible service. \hl{To ensure the quality of the data sets, a full manual review was done in January, after May 25, and in October 2018.} For the measurements in the months in between, we used the lists from previous months to download privacy policies.
In the majority of cases, we found links to privacy policies in the footer of a website (an approach also used by Libert~\cite{libert_2018}) or through links in cookie consent notices. When there was no footer or no link to a privacy policy, annotators inspected the site in more detail. Several websites made it rather complicated for users to find these links as they, for example, had a privacy policy link in the site's footer but used \textit{infinite scrolling} to dynamically add more content when the user scrolled to the bottom of the page, moving the footer out of the visible area again. Sites without footers were inspected for links to other documents that may contain information about the handling of personal data like terms of service, user agreements, legal disclaimers, contact forms, registration forms, or imprints.

\vspace{-1ex}
\subsection{Archival data}
The GDPR was passed in April 2016, allowing for a two-year grace period before it went into effect. Given that we started collecting data in January 2018, we used the Internet Archive's Wayback Machine
to retrieve previous versions of the privacy policies in our dataset. This allowed us to analyze whether and when privacy policies had been changed before our data collection started. Using the Wayback Machine's API, we requested versions for each policy URL for each month between March 2016 and December 2017. On average, we were able to access previous versions for 2,187 policies for each month. The extent of this dataset is limited due to the fact that not every website or page is archived by the Internet Archive and some of the pages we tried to access might not have existed previously.

\subsection{Data Cleaning}
After retrieving a total of 112,041 privacy policies, we pre-processed these files with Boilerpipe, an HTML text extraction library, to remove unnecessary HTML code from the documents \cite{boilerplate_2010}. Boilerpipe removes HTML tags and identifies the main text of a website removing menus, footers, and other additional content. We validated the results with text that had been manually selected while inspecting sites for privacy policies. Except for policies that were very short (less than four sentences) and excluded because Boilerpipe was not able to identify their main text, it correctly extracted the policy texts.
We scanned the remaining files for error messages in multiple languages and manually inspected sentences many texts had in common to exclude those if they indicated an error.

We observed some websites that linked to a privacy policy at a domain different from its own, either as the only privacy policy link or in addition to the website's own policy. A valid and common reason for a privacy policy being linked from multiple hosts was \hl{websites referencing} the policy of a parent company, \eg, RTL Group (linked on 11 domains), Gazeta.pl (9), Vox Media Group (4). We excluded these (duplicate) policies from further analysis. \hl{We also marked as offline websites linking to privacy policies of unrelated third parties (\eg, Google or domain grabbing services) as they evidently did not have a policy specific to their data collection practices.}

72 sites used JavaScript to display their privacy policies, which was not properly detected by our script, resulting in file downloads that contained the websites' home pages instead of their privacy policies. Unfortunately, we did not discover this issue until

the analysis, at which point we decided to exclude them. We also had to exclude 163 websites from our content analysis that provided their policies as a file download (\eg, as a PDF or DOC file) -- although their availability was detected, our crawler was not designed to process these.

After the data cleaning process, our dataset for text mining consisted of 81,617 policies from 9,461 different URLs and 7,812 domains.
We also removed lines from the files downloaded from the Internet Archive that contained additional information about the data source.

To compare different versions of policies and policies from different websites we used the Jaccard similarity index on a sentence level~\cite{huang2008similarity}, which is commonly used to identify plagiarism \cite{leskovec_mining_2014}. The Jaccard index measures similarity as the sum of the intersection divided by the sum of the union of the sentences. It ranges between 0 and 1, where 1 means two documents only have the same sentences.

We used the Polyglot\footnote{https://github.com/aboSamoor/polyglot.} library to split the texts into sentences and stored a policy as a list of MD5-hashed sentences to speed up the text comparison process. This resulted in a database of policies where each policy consisted of a number of hashed sentences \(P_{domain,url,crawl}=[h_1,h_2,..h_n]\) and calculated the similarity \(S\) between two policies \(P_x\) and \(P_y\) where $x$ and $y$ marked documents from two different crawls but from the same domain and URL as
\[
S(P_x, P_y)  = \frac{P_x \cap P_y}{ P_x \cup P_y}.
\]
We compared monthly versions of each crawl to analyze when and if privacy policies had changed. We also compared versions over larger intervals, \eg, between January 2017 and December 2017. To do the latter, we had to exclude several websites from the comparison, \eg, when there was no data available on the Internet Archive but also when the URL of their privacy policy had changed. Although we downloaded pages that appeared with new links, we only compared texts from the same URLs as we were not able to automatically determine which version to compare.

For example, multiple websites previously listed their privacy policy as part of the terms of service page and then moved it to a separate page. Again, we took a conservative approach and only compared different versions of the same files.

The Jaccard index would still detect a change compared to the first document we had on file, in that case, the terms of service.

Lastly, we applied lemmatization/stemming to the documents to perform an analysis on the word level and check whether privacy policies mentioned phrases specific to the GDPR. First, we created a word list with translations of important phrases from Articles 6 and 13 GDPR. The EU provides official translations of all documents in 24 different languages from which we extracted the corresponding phrases.

Leveraging our extended personal networks, we recruited native speakers for 17 of the 24 languages to check and validate the word lists.\footnote{We could not find native speakers for Danish, Latvian and Lithuanian but did our best to validate the words using dictionaries and translation tools.}

We then searched for these words by first determining the language of a policy using two libraries, The Language Detection Library \footnote{https://github.com/shuyo/language-detection.} and Polyglot.

We excluded 1.7\,\% of texts from our analysis because the libraries produced diverging results.

Because of the high diversity in the policies' languages -- 24 official languages of EU member states, plus 7 other languages occurring in our dataset

-- we used three different natural language processing libraries (NLTK, Spacy, and Polyglot) to process the policies and compared the results to ensure that the linguistic properties of the respective languages such as conjugation where factored in correctly. We chose Polyglot as it performed best on the specific word lists we had created.

Since Polyglot does not include lemmatization, we utilized distinct lemmatization lists.\footnote{Available at \url{https://github.com/michmech/lemmatization-lists}}. We also utilized Named Entity Recognition (NER) and regular expressions as an ensemble approach to search the policies for contact data.

\subsection{Limitations}
\label{sec:limitations}

Scheitle et al.~\cite{scheitle_toplists_2018} showed that many publicly available top lists, including Alexa, are biased, fluctuate highly, and that there are substantial differences among lists.

Indeed, we observed high fluctuation as, on average, a country's top list from January and May only had 387 entries in common. Nevertheless, we relied on Alexa's top lists, as they are the only source for country-specific rankings. However, we accounted for high fluctuation by refraining from analyzing correlations between the top list ranking and other factors measured, except for the impact of consent notice libraries.

We accounted for bias potentially introduced due to the rankings used by

conducting the pre-post analysis only on domains present in the January top list. To account for potential top list manipulation~\cite{lepochat_alexa_2018},

especially give some countries' small population,

we excluded domains that were offline during one of the crawls or were blocked by the protection mechanisms of the browser. Moreover, the obligation to comply with legal regulations is independent of the legitimacy of being listed in top lists.

Regarding the use of GDPR-related terms in text analysis, our keyword list can only provide limited insights into the GDPR compliance of policy texts. Although we created a comprehensive list of translations of relevant terms, privacy policies are not required to use these terms. In fact, the GDPR's requirement to provide privacy policies in an ``intelligible'' form could potentially decrease the use of legal jargon in privacy policies, although we did not see evidence of that in our dataset. Nevertheless, our keyword lists should be seen as a starting point for additional research and analysis in order to assess legal compliance in more detail and at scale.

\section{Evaluation of Privacy Policies}
\label{sec:evaluation}
In total, the lists of the 500 most frequently visited websites for all 28 EU member states in January 2018 contained 6,759 different domains; the \hl{final list in November contained 13,458 domains}. Unless mentioned otherwise the pre-/post-GDPR comparison is based on the data points for the domains first annotated in January, while the analysis of the cookie consent notices is based on the extended list we had created by the end of May. The overall prevalence of privacy policies on these websites was already high (79.6\,\%) before the GDPR went into effect and only increased slightly to 84.5\,\% afterwards. However, we found big differences among the 28 EU member states, with privacy policy rates between 75.6\,\% and 97.3\,\% at the end of May, and also between different content categories varying from 53.7\,\% and 98.2\,\%. Although the GDPR was officially adopted in 2016, half of the websites (50.4\,\%) updated their privacy policies in the weeks before May 25, 2018. 15\,\% did not make any update since the adoption.

The GDPR's most notable (and visible for users) effect we observed is the increase of cookie consent notifications, which rose from 46.1\,\% in January to 62.1\,\% in May.
We found that especially popular websites implement cookie consent notices and choices using third party libraries.
Our in-depth analysis of

common libraries found in our dataset revealed shortcomings in how those consent mechanisms can satisfy the requirements of Article 6 of the GDPR (see Section~\ref{sec:cbtypes} for details).

\subsection{Privacy Policies}
\label{subsec:pp}

Our dataset of privacy policies was based on 6,759 domains since multiple services (e.g., Facebook and Google) appear in more than one country's top list. Of those domains, 5,091 had a complete or partial privacy policy statement. In January, our system found the majority of policies (3,476) automatically, the remaining 3,283 sites were checked manually, resulting in the identification of another 1,624 privacy policies. 1,276 websites did not have a privacy policy and the remaining 383 websites could not be reached.

\subsubsection{Websites added policies}
Table \ref{table:ppstats} gives an overview of the changes in the number of websites with privacy policies for the (a) 500 most popular websites in a country and (b) country-specific top-level domains (TLD). For this analysis, we compared the results of January 2018 with those from right after May 25, 2018. In both sets, we excluded sites that we found to be offline during at least one of the crawls.
\hl{Results for October 2018 only slightly deviate from the measurement made at end of May. The average increase from May to October was +1.0 percentage point.}

The data shows that the majority of websites (79.6\,\%) already had privacy policies in January 2018. That level rose by 4.9\,\% to 84.5\,\% after May 25, 2018. However, there are clear differences in the country and domain level. Countries with a lower rate of privacy policies added more privacy policies than those where privacy policies were already common. For example, in Latvia's top-500 list 10.2\% of the websites added privacy policies, and an even higher amount (+27\,\%) of all websites with the Latvian TLD .lv added one. At the same time, in countries like Spain (ES), Germany (DE) or Italy (IT), where over 90\,\% of websites on the top lists had privacy policies, few sites added them. On the domain level, these few additional sites helped to reach 100\,\%.

We also checked the prevalence of privacy policies on non-EU and generic TLDs, of which we found 207 unique ones in our dataset; 39 occurred in the top lists of 20 or more countries. Table~\ref{table:ppstats} lists the 5 most frequently found TLDs that are not EU-country specific. Besides generic TLDs (.com, .org, .info, .net, .eu, .tv) Russia's TLD .ru frequently showed up in top lists of countries with a Russian-speaking minority.

Table~\ref{table:categoriepp} shows data from the same comparison between January and May ordered by website category. Overall, 4.9\,\% of websites added policies, note that the average differs since websites were listed in multiple top lists and could also be assigned multiple categories. Based on these findings, GDPR seems to have had the biggest impact on sites that are more likely to collect sensitive information like health or sports-related websites or that are connected to children (Kids \& Teens, Education). The processing of the personal information of children must also adhere to higher standards in the GDPR.

It is a positive result that the highest rates of privacy policies occur in the Finance, Shopping, and Health categories, where websites routinely process more sensitive data.
\hl{Between May and October, 10 sites removed their privacy policy. The manual analysis showed that in most cases the sites were redesigned and no policy was (re-)added. For some websites, \eg, Feedly.com, the privacy policy was still available under a link we had previously stored, but the link is not made available to users that are not already registered with the service.}
In general, more websites added policies when they had been less prevalent in their country/category. \claim{The largest changes were observed in the Baltic states (on .lv, .lt and, .ee domains), but affected all top lists.}

\begin{table}

	\centering
	\caption{Availability of privacy policies in the top 500 websites by country, pre- (January 2018) and post-GDPR (after May 25, 2018).}

\begin{tabular}{p{.15cm}p{.15cm}p{.6cm}p{.6cm}p{.8cm}p{.15cm}p{.15cm}p{.6cm}p{.6cm}p{.8cm}}\\

	\toprule
	 & \multicolumn{4}{c}{top list} & \multicolumn{5}{c}{TLD} \\
	 & \multicolumn{1}{c}{\textbf{N}}
	 & \multicolumn{1}{c}{\textbf{Pre}}
	 & \multicolumn{1}{c}{\textbf{Post}}
	 & \multicolumn{1}{c}{\textbf{Diff}}
	 &
	 & \multicolumn{1}{c}{\textbf{N}}
	 & \multicolumn{1}{c}{\textbf{Pre}}
	 & \multicolumn{1}{c}{\textbf{Post}}
	 & \multicolumn{1}{c}{\textbf{Diff}}  \\

	\midrule

AT & 455 & 91.6\,\% & 94.5\,\% & 2.9\,\% & .at & 132 & 95.5\,\% & 98.5\,\% & 3.0\,\%\\
BE & 460 & 89.6\,\% & 92.4\,\% & 2.8\,\% & .be & 141 & 92.2\,\% & 97.9\,\% & 5.7\,\%\\
BG & 451 & 83.1\,\% & 88.9\,\% & 5.8\,\% & .bg & 166 & 80.1\,\% & 89.8\,\% & 9.6\,\%\\
CY & 432 & 76.4\,\% & 83.6\,\% & 7.2\,\% & .cy & 58 & 62.1\,\% & 69.0\,\% & 6.9\,\%\\
CZ & 459 & 81.9\,\% & 88.0\,\% & 6.1\,\% & .cz & 251 & 80.9\,\% & 89.2\,\% & 8.4\,\%\\
DK & 447 & 91.3\,\% & 95.1\,\% & 3.8\,\% & .dk & 174 & 95.4\,\% & 99.4\,\% & 4.0\,\%\\
DE & 455 & 88.8\,\% & 91.6\,\% & 2.9\,\% & .de & 172 & 98.8\,\% & 100.0\,\% & 1.2\,\%\\
EE & 441 & 63.5\,\% & 76.2\,\% & 12.7\,\% & .ee & 132 & 56.8\,\% & 72.7\,\% & 15.9\,\%\\
ES & 429 & 90.0\,\% & 92.1\,\% & 2.1\,\% & .es & 86 & 98.8\,\% & 100.0\,\% & 1.2\,\%\\
FI & 462 & 85.1\,\% & 92.0\,\% & 6.9\,\% & .fi & 145 & 80.7\,\% & 93.1\,\% & 12.4\,\%\\
FR & 453 & 90.7\,\% & 93.6\,\% & 2.9\,\% & .fr & 139 & 95.7\,\% & 98.6\,\% & 2.9\,\%\\
GB & 463 & 95.5\,\% & 97.2\,\% & 1.7\,\% & .uk & 108 & 98.1\,\% & 98.1\,\% & 0.0\,\%\\
GR & 443 & 77.9\,\% & 83.7\,\% & 5.9\,\% & .gr & 233 & 72.1\,\% & 80.3\,\% & 8.2\,\%\\
IE & 447 & 91.1\,\% & 93.1\,\% & 2.0\,\% & .ie & 104 & 98.1\,\% & 99.0\,\% & 1.0\,\%\\
IT & 423 & 90.3\,\% & 93.9\,\% & 3.5\,\% & .it & 174 & 96.6\,\% & 97.7\,\% & 1.1\,\%\\
HU & 440 & 85.7\,\% & 90.5\,\% & 4.8\,\% & .hu & 228 & 85.5\,\% & 91.2\,\% & 5.7\,\%\\
HR & 430 & 82.8\,\% & 86.3\,\% & 3.5\,\% & .hr & 141 & 82.3\,\% & 84.4\,\% & 2.1\,\%\\
LV & 434 & 59.9\,\% & 75.6\,\% & 15.7\,\% & .lv & 126 & 46.8\,\% & 73.8\,\% & 27.0\,\%\\
LT & 452 & 67.9\,\% & 78.1\,\% & 10.2\,\% & .lt & 174 & 58.0\,\% & 73.6\,\% & 15.5\,\%\\
LU & 440 & 81.4\,\% & 84.8\,\% & 3.4\,\% & .lu & 61 & 65.6\,\% & 73.8\,\% & 8.2\,\%\\
MT & 446 & 86.3\,\% & 88.3\,\% & 2.0\,\% & .mt & 46 & 63.0\,\% & 71.7\,\% & 8.7\,\%\\
NL & 459 & 86.3\,\% & 90.0\,\% & 3.7\,\% & .nl & 115 & 96.5\,\% & 100.0\,\% & 3.5\,\%\\
PL & 462 & 91.1\,\% & 94.4\,\% & 3.2\,\% & .pl & 256 & 93.4\,\% & 96.5\,\% & 3.1\,\%\\
PT & 430 & 85.6\,\% & 88.6\,\% & 3.0\,\% & .pt & 116 & 86.2\,\% & 91.4\,\% & 5.2\,\%\\
RO & 434 & 81.3\,\% & 85.9\,\% & 4.6\,\% & .ro & 160 & 86.3\,\% & 91.9\,\% & 5.6\,\%\\
SE & 459 & 89.1\,\% & 93.2\,\% & 4.1\,\% & .se & 166 & 87.3\,\% & 94.6\,\% & 7.2\,\%\\
SK & 438 & 79.5\,\% & 86.3\,\% & 6.8\,\% & .sk & 189 & 73.5\,\% & 84.1\,\% & 10.6\,\%\\
SI & 451 & 91.4\,\% & 95.6\,\% & 4.2\,\% & .si & 132 & 90.9\,\% & 96.2\,\% & 5.3\,\%\\

	\midrule
Total & 6357 & 79.6\,\% & 84.5\,\% & 4.9\% &  & 4125 & 82.7\,\% & 89.4\,\% & 5.7\,\%\\
\midrule
& & & & & .com & 2026 & 82.5\,\% & 83.9\,\% &1.4\,\% \\
& & & & & .ru & 147 & 65.6\,\% & 68.8\,\% & 3.2\,\%\\
& & & & & .org & 122 & 47.5\,\% & 50.0\,\% & 2.5\,\%\\
& & & & & .net & 248 & 64.6\,\% & 70.6\,\% & 6.0\,\%\\
& & & & & .eu & 43 & 58.1\,\% & 67.4\,\% & 9.3\,\%\\

	\bottomrule
	\label{table:ppstats}
\end{tabular}
\end{table}

\begin{table}

	\centering
	\caption{Availability of privacy policies per website category, pre- (January 2018) and post-GDPR (after May 25, 2018).}

\begin{tabular}{p{3cm}p{.6cm}p{.6cm}p{.6cm}p{.8cm}}\\
	\toprule
	 \multicolumn{1}{c}{Category}
	 & \multicolumn{1}{c}{\textbf{n}}
	 & \multicolumn{1}{c}{\textbf{pre}}
	 & \multicolumn{1}{c}{\textbf{post}}
	 & \multicolumn{1}{c}{\textbf{diff}}
 \\
	\midrule

Adult & 256 & 68.8\,\% & 72.7\% & 3.9\%\\
Arts \& Entertainment & 521 & 70.1\,\% & 75.8\,\% & 5.7\,\%\\
Business & 529 & 81.5\,\% & 87.3\,\% & 5.8\,\%\\
Computers & 686 & 87.9\,\% & 90.8\,\% & 2.9\,\%\\
Education & 380 & 70.0\,\% & 79.7\,\% & 9.7\,\%\\
Finance & 427 & 92.3\,\% & 96.5\,\% & 4.2\,\%\\
Games & 245 & 87.8\,\% & 92.7\,\% & 4.9\%\\
Government & 132 & 66.7\,\% & 73.5\,\% & 6.8\,\%\\
Health & 99 & 89.9\,\% & 97.0\,\% & 7.1\,\%\\
Home & 134 & 97.8\,\% & 99.3\,\% & 1.5\,\%\\
Kids and Teens & 37 & 83.78\% & 91.89\% & 8.11\%\\
News & 958 & 80.8\,\% & 86.6\,\% & 5.8\,\%\\
Recreation & 90 & 81.1\,\% & 86.7\,\% & 5.6\,\%\\
Reference & 497 & 83.5\,\% & 88.1\,\% & 4.6\,\%\\
Regional & 108 & 81.5\,\% & 88.0\,\% & 6.\,\%\\
Science & 31 & 90.3\,\% & 96.8\,\% & 6.5\,\%\\
Shopping & 925 & 94.4\,\% & 98.2\,\% & 3.8\,\%\\
Society \& Lifestyle & 444 & 86.0\,\% & 90.1\,\% & 4.1\,\%\\
Sports & 267 & 80.2\,\% & 86.5\,\% & 6.3\,\%\\
Streaming & 337 & 50.5\,\% & 53.7\,\% & 3.2\,\%\\
Travel & 250 & 88.8\,\% & 93.2\,\% & 4.4\,\%\\
 	\midrule
avg.& 350.14 & 86.9\,\% & 5.3\,\% & 5.4\,\%\\

	\bottomrule
	\label{table:categoriepp}
\end{tabular}
\vspace{-0.6cm}
\end{table}

\subsubsection{Changes in privacy policies}

\begin{figure}
	\centering
	\includegraphics[width=.45\textwidth]{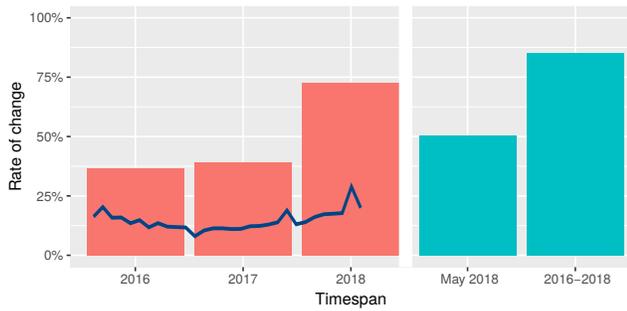}
	\caption{Percentage of policies changed in a certain time span. n(2016) = 860, n(2017) = 806, n(2018) = 726, n(May2018) = 6195, n(2016-2018) = 1610. The line shows the average month-to-month change.}
	\label{fig:change_rate}
	\vspace{-0.8cm}
\end{figure}

We compared different versions of privacy policies to see if they changed and whether these changes were GDPR-related.
The majority of websites updated their privacy policies in the last two years. Comparing versions from March 2017 (before the GDPR was passed) and May 2018,  85.1\,\% were changed at least once. About 72.6\,\% of those policies were (also) updated between January and June 2018, but the majority of changes (50.0\,\%) occurred within one month preceding May 25. \hl{Analyzing the variance between two month using ANOVA showed significant changes from November to December 2017 (most likely due to the fact that policies before that date were based on archival data) and around the GDPR deadline early May to June to July.}

Some websites seemingly missed the GDPR deadline: 118 sites that had not updated their privacy policy since early 2016 did so between our two post-GDPR measurements at the end of May and the end of June 2018.

In all cases, privacy policy changes meant the addition of text to the privacy policy. The average text length rose from a mean of 2,145 words in March 2016 to 3,044 words in March 2018 (+41 percentage points in 2 years) and increased another 18 percentage points until late May (3,603 words).\footnote{We refrained from comparing policy lengths across countries due to language differences impacting length (\eg, the use of compounds instead of separate words).} This demonstrates a tension between the GDPR's requirement for concise and readable notices with its additional disclosure requirements, such as mentioning the legal rights of a data subject, providing the data processor's contact information, and naming its data protection officer.

\subsubsection{GDPR compliance issues}

By the end of May, 350 of the 1,281 websites that did not have a policy in January had added one.

The remaining 931 sites can be considered not compliant with the GDPR's transparency requirements due to the lack of a privacy policy or similar disclosure.

Websites without privacy policy remain most common in the Baltic states. More than 24\% of top-listed sites in Lithuania, Latvia, and Estonia still had no privacy policy. While some of those pages might not be actively maintained or may not care about legal obligations due to illicit content, 73 websites have no privacy policy but serve a cookie consent notice (down from 161 in January). We even found 14 websites that added this kind of notification in 2018 without adding a privacy policy.

\subsubsection{Policy content}

Comparing the content of privacy policies between January and May, we saw that an additional 9\,\% of policies contained e-mail addresses, up from 37.7 to 46.6\,\%. Similarly, an additional 9\,\% mentioned a data protection officer.
Searching for GDPR keywords in our set of policies in all languages yielded an increase in the use of all keywords.
Since website owners are not required to use these specific terms (see \ref{sec:limitations}), we focused on analyzing the change in their importance by ranking the terms based on the number of policies that included them. Overall, terminology related to user rights (``erasure'' (+8\,\%), ``complaint'' (+11\,\%), ``rectification'' (+6\,\%), ``data portability''(+7\,\%)) appeared more often.
We also saw an increase in mentions of possible legal bases of processing. While the number of policies mentioning consent was stable (J: 28\,\%, M: 29.2\,\%), an increasing number of policies explicitly mentioned other aspects described in Article 6 GDPR like ``legitimate interest'' (J: 7\,\%, M: 19.2\,\%).

\subsubsection{Tracking and cookies}

We did not observe a significant change in the use of tracking services or cookies. In January, websites used on average 3.5 third-party tracking services that would be blocked by an off-the-shelf ad blocker.

Still, some websites made notable changes: we manually checked websites that did not use trackers in June but did so in January and found that 146 stopped using ad or tracking services and 37 did not track before explicit user consent was given. Notable examples are \url{washingtonpost.com} and \url{forbes.com}. Only after consenting into tracking -- or subscribing to paid services -- users are directed to the regular homepage of these sites.

In May, right before the GDPR came into effect, and in June we measured the number of first- and third-party cookies a website sets by default. Regarding third-party cookies no effect is visible; websites set about 5.4 cookies on average. The number of first-party cookies decreased from 22.2 to 17.9 cookies on average. This effect can be explained by a decrease in first-party cookie use in Croatia (-11.3) and Romania (-21.1). The medians stayed the same for both cookie groups.

\subsubsection{HTTPS}

We also measured whether the adoption of HTTPS by default changed over the course of \hl{twelve} months. We always checked the HTTP address of a host and observed whether the visited website automatically redirected to HTTPS. Our data confirm a general trend towards HTTPS that was reported before~\cite{felt_measuring_2017}. Figure \ref{fig:https} shows the increase in the use of HTTPS by default from 59.9\,\% in December 2017 to 80.2\,\% in November 2018.
\hl{At the end of May, 70.8\,\% of websites redirected to HTTPS, close to the 74.7\,\% reported by Scheitle et al.~\cite{scheitle_toplists_2018}, who measured the HTTPS capabilities of the Alexa top 1 million websites.}
\hl{The average increase was +1.9 percentage points in a month-by-month comparison. Statistically significant changes in the variance (ANOVA) were found from December 2017 to January 2018 (+2.9), early May to June (+3.9), and October to November 2018 (+2.7). The high increase from May to June was preceded and followed by months of less increase, which can be interpreted as a concentration of activities around the GDPR enforcement date that followed an overall trend.}
\hl{Looking at the TLD level, the majority (18 out of 28) show an adoption larger than 80\,\% in November 2018. For three countries, we found an increase of more than 30 percentage points (.pl, .gr., .es), but only for .es the adoption is now above the average.}
\begin{figure}
	\centering
	\includegraphics[width=.45\textwidth]{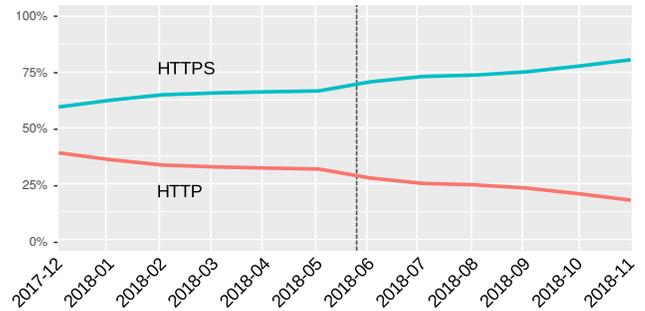}
	\caption{\hl{Change in HTTPS adoption over time. The dotted line marks the GDPR enforcement date.}}
	\label{fig:https}
	\vspace{-0.8cm}
\end{figure}

\claim{Our findings indicate that \fixme{at the time the GDPR came into force the number of websites with privacy policies increased}, affecting some countries and sectors more than others.} Effects have so far been limited to transparency mechanisms as the use of tracking and cookies appears largely unchanged. In the next sections, we focus on a second development, the increase in the use of cookie consent notices, which, in principle, should not only inform users but also offer actual choice.

\section{Studying Cookie Consent Notices}
\label{sec:cbtypes}

In January and May, we manually inspected all websites for cookie consent notices. In January, we only noted whether a website displayed a cookie banner or not. Because the observed sophistication of cookie banners increased substantially, during the May annotation, we also analyzed and categorized the type of consent notice based on its interaction options.
We identified the following distinct types with examples shown in Figure~\ref{fig:cookiebartypes}:

\begin{figure*}
	\centering
	\includegraphics[width=1.0\textwidth]{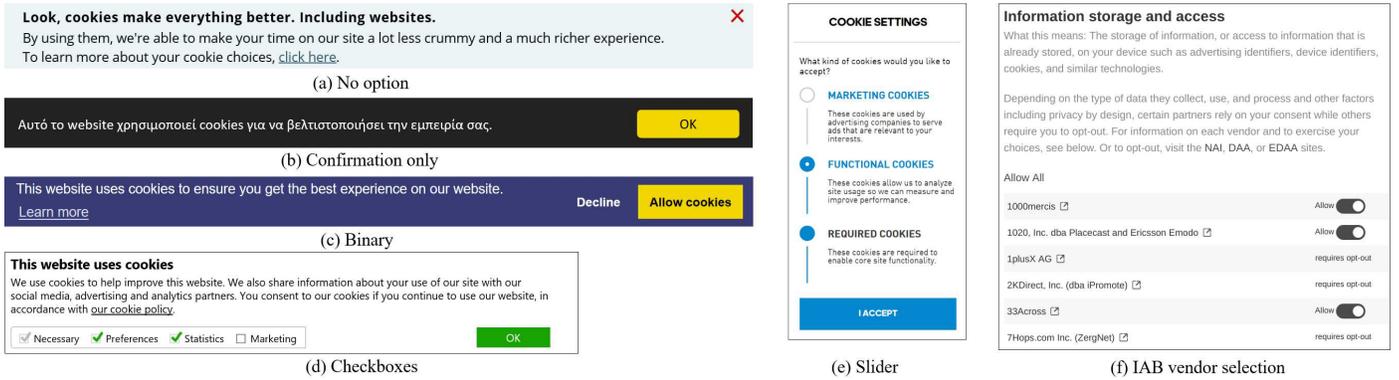}
	\caption{Types of cookie consent notices with different interaction models.}
	\label{fig:cookiebartypes}
	\vspace{-0.6cm}
\end{figure*}

\textbf{No Option:} Cookie consent notices with \emph{no option} (Figure~\ref{fig:cookiebartypes} (a))
simply inform users about the site's use of cookies. Users cannot explicitly consent to or deny cookie use. This category also includes banners that feature a clickable button whose label cannot be considered to express agreement (\eg, ``Dismiss,'' ``Close,'' or just an ``X'' to discard the banner).

\textbf{Confirmation:} In contrast, \emph{confirmation-only} banners (Figure~\ref{fig:cookiebartypes} (b)) feature a button with an affirmative text such as ``OK'' or ``I agree''/``I accept'' which can be understood to express the user's consent.

\textbf{Binary} consent notices (Figure~\ref{fig:cookiebartypes} (c)) give users the options to explicitly agree to or decline all the website's cookies.

\textbf{Slider:}  More fine-grained control is offered by cookie banners that group the website's cookies into categories, mostly by purpose. \emph{Slider}-based notices (Figure~\ref{fig:cookiebartypes} (d)) arrange these categories into a hierarchy. The user can move a slider to select the level of cookie usage they are comfortable with, which implies consent with all the previously listed categories.

\textbf{Checkbox}-based notices (Figure~\ref{fig:cookiebartypes} (e)) allow users to accept or deny each category individually. The number of categories varied, ranging from 2 to 10 categories; we observed that most notices of the ``checkbox'' type featured 3--4 different cookie categories. A common set of categories comprises advertising cookies, website analytics, personalization, and what is usually referred to as (strictly) necessary cookies, such as shopping cart cookies. According to Article~5(3) of the ePrivacy Directive (2002/58/EC), this type of cookies does not require explicit user consent.

\textbf{Vendor:} \hl{We assigned this category to banners that allow users to toggle the use of cookies for each third party individually. Figure~\ref{fig:cookiebartypes} (f) shows one such mechanism.}

\textbf{Other:} \hl{This category, assigned five times in total, was used for cookie banners that did not match any other category, \eg, one site allowed users to choose between two ``cookie profiles''.}

In addition to the cookie banner annotation, all websites were manually categorized by topic to specify what information or services they provide. We used Alexa's website categorization scheme.\footnote{\url{https://www.alexa.com/topsites/category}} but performed the categorization manually since Alexa only provided categories for about a third of the websites in our data set. We also added the categories ``Government'' and ``Streaming'' because our dataset contained a substantial number of websites fitting those categories.

\subsection{Analysis of Cookie Consent Libraries}

During manual website annotation,
we noticed that websites made use of third-party implementations to provide cookie consent notices.
This raised questions about how common certain cookie consent solutions are and to what degree they can help website owners comply with Directive 2002/58/EC and the GDPR.

\hl{We compiled a list of the cookie consent libraries identified during manual annotation.} If possible, we downloaded each library or requested access to a (demo) account from the vendor. We subsequently implemented each consent solution -- one at a time -- into a live WordPress website. We then visited the site using Microsoft Edge 41 configured to not block any cookies, interacted with the cookie banner, and used Edge's Developer Console to observe the effect of user selection on the cookies stored to the machine.
For each library, we tested the user interfaces it offered and whether its settings and documentation allowed us to block and unblock cookies (i.e., we did not write any custom code to implement new core functionality). We also tested if the libraries provided mechanisms to reconsider a previous consent decision and to log and store the users' consent, as required by Article 7 GDPR.

It is in the interest of web service providers not to display consent notices to users that are not subject to GDPR. Thus, many libraries offer the option to display the notice only to users accessing the site from specific regions of the world. We tested these geolocation features using Tor Browser and a circuit

exiting in a country for which the cookie banner was configured not to show up.

We measured the popularity of  identified cookie libraries in a separate scan of domains' home pages in July \hl{and December} 2018.

To determine if a website used a cookie library, we reviewed the default locations of JS and CSS resources and likely variants based on the installation instructions. Additionally, we checked for requests to third parties used by the libraries.

We manually verified this procedure with a list compiled during the manual annotation phase. To reflect the exposure a library or service has to end users, we calculated a score based on the ranking of the domain in Alexa.com's EU top lists. This favors domains which are highly ranked in many top lists over domains which are only in a single top list.

This better accounts for the exposure a library has to end users. This $Score$ inherits the bias the Alexa top list has (see Section \ref{sec:limitations}).
It is calculated by subtracting the $Rank_{toplist,i}$ of a domain from $501$ for each top list ($N$) and summing up these values. Sites no longer present in the top lists were assigned rank $501$. The $Score$ is then normalized by dividing by $N$: $$Score=\frac{\sum_{i=1}^{N}{501-Rank_{toplist,i}}}{N}$$

\subsection{Limitations}
Parts of our study were conducted with automated browsers using a server hosted on a known server farm. It is known that some websites change their behavior when an automated browser or specific server IP addresses are detected. We observed that several websites using Cloudflare's services blocked direct requests and asked to resolve a CAPTCHA before redirecting to the actual site. As described above, we checked for these effects as we manually visited all websites to determine, \eg, which type of cookie banner they used.
Another drawback of

our technical setup was that some websites might have changed their default language based on the IP of the server (in Germany) or the default browser language (English). While this might have influenced the language of the privacy policy and cookie banner presented, it should not have changed the fact that either exists.

\begin{figure*}[b]

    \subfloat[\hl{Cookie banner types by country (October 2018). Dotted line indicates the average.}]{{\includegraphics[width=0.6\textwidth]{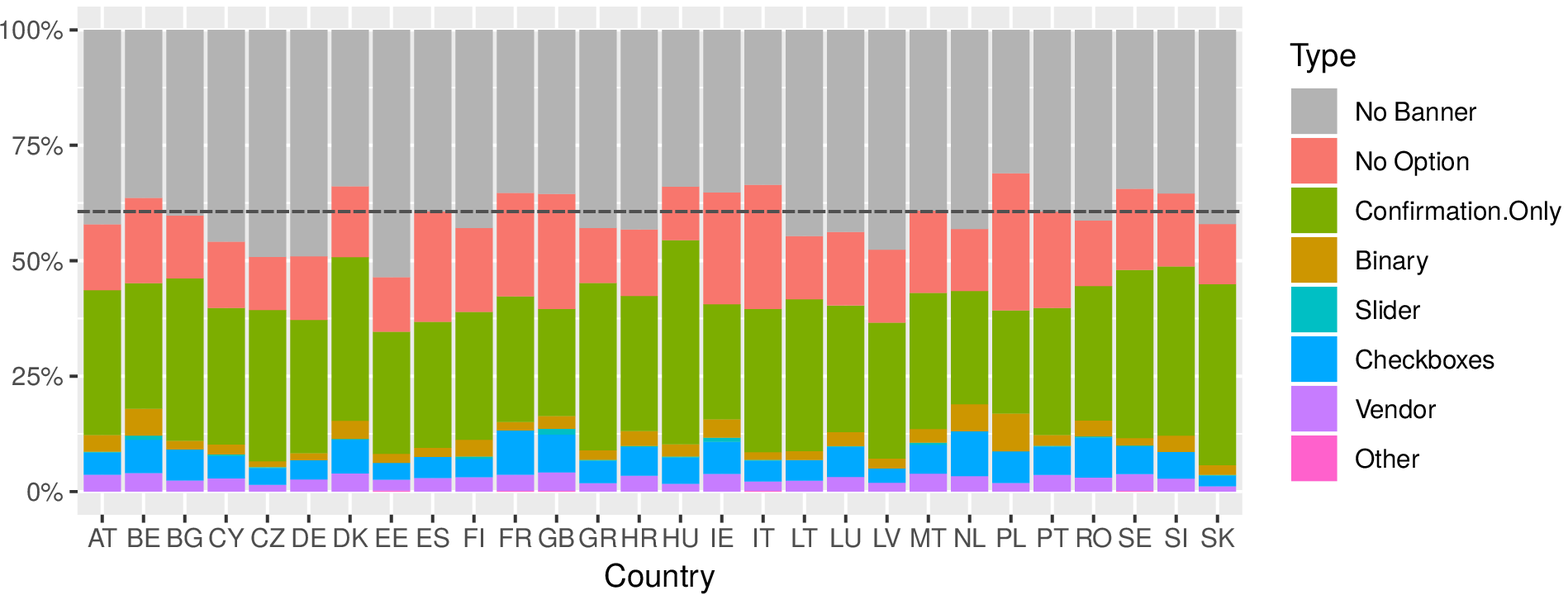} }}
    \qquad
    \subfloat[Distribution of cookie banner libraries based on the websites' Alexa rank \hl{(December 2018).}]{{\includegraphics[width=0.3\textwidth]{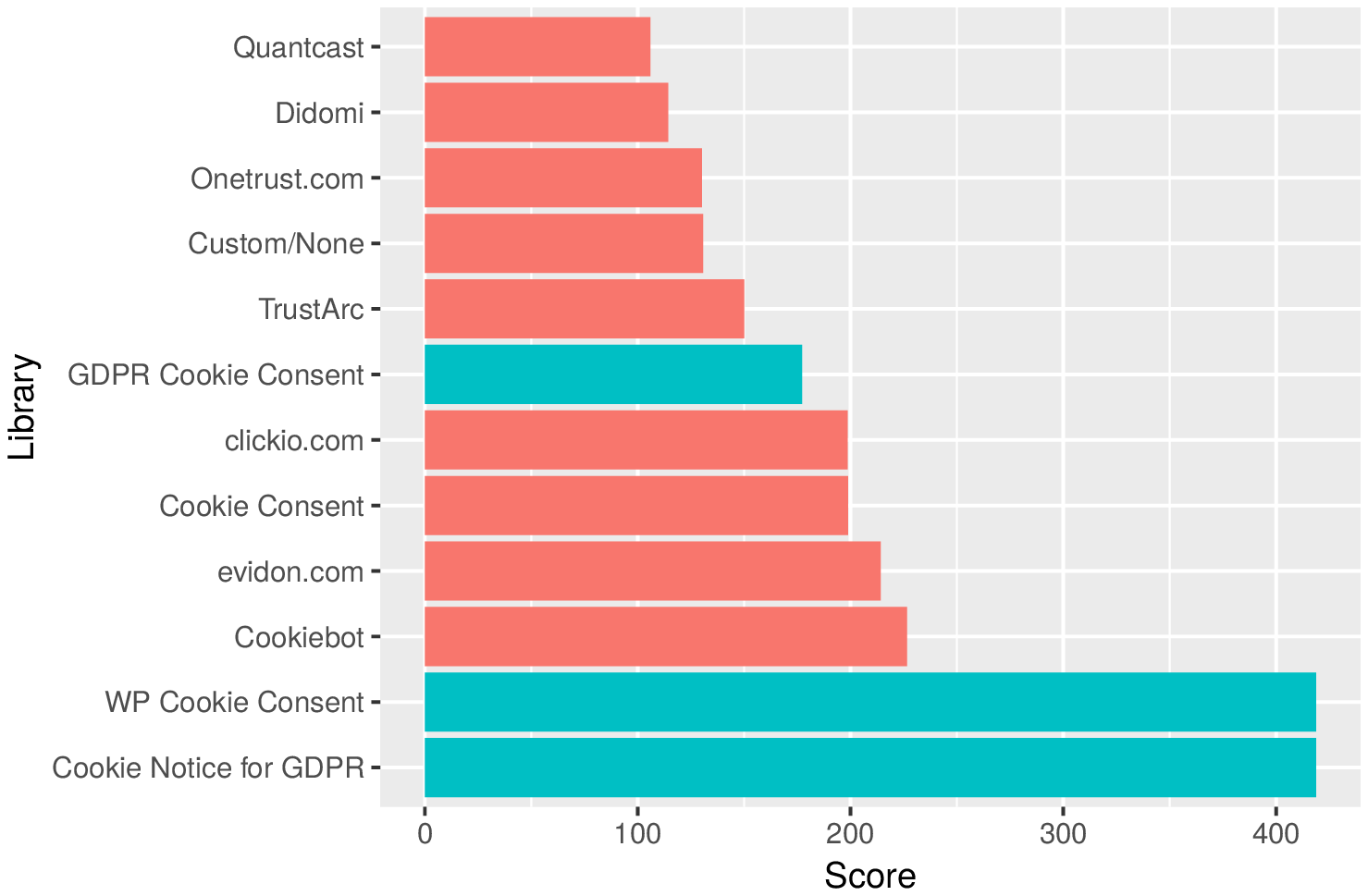} }}
\caption{Distribution of cookie consent notices and popularity of libraries.}
\label{fig:cookiebanner_distribution}
\label{fig:cookiebanner_distribution_alexa}
\vspace{-0.7cm}
\end{figure*}

\section{Evaluation of Cookie Consent Notices}
\label{sec:cookieeval}

\hl{We found that the adoption of cookie consent notices had increased across Europe, from 46.1\,\% in January to 62.1\,\% at the end of May (post-GDPR) and reached 63.2\,\% in October 2018}. Adoption rates significantly differ across individual member states, as does the distribution of different types of consent notices. The libraries we encountered on popular sites do not always support important features to fulfill GDPR requirements like purpose-based selection of cookies and consent withdrawal.

\subsection{Adoption}
\label{sec:cb-adoption}

Table~\ref{table:cookiebardevstats} compares the prevalence of cookie consent notices in January 2018 with May 2018. Grouped by Alexa country list, the percentage of sites featuring a consent notice, on average, has increased, ranging from +20.2 percentage points in Slovenia to +45.4 in Italy. Looking at the sites by top-level domain (TLD), the average adoption rate increased from 50.3\,\% to 69.9\,\% post-GDPR. For the .nl and .si TLDs, the number of sites implementing a cookie banner did not increase substantially from January to May 2018 as they both already had high adoption rates of 85.2\,\% and 75.8\,\%, respectively. The highest increase in cookie banner prevalence by TLD was observed in Ireland -- for the 104 .ie domains in our dataset, the adoption rate increased from 17.3\,\% to 87.5\,\%.

Figure~\ref{fig:cookiebanner_distribution_alexa} (a) shows the distribution of the different types of cookie consent notices (see Section~\ref{sec:cbtypes}) by country post-GDPR (end of May 2018). The use of \textit{checkbox}-based cookie consent notices stands out in France and Slovenia, while websites in Poland use the highest number of \textit{no-option} notices.

\begin{table}
	\centering
	\caption{Availability of cookie consent notices in the top 500 websites by country, pre- (January 2018) and post-GDPR (after May 25, 2018).}
	\label{table:cookiebardevstats}
\begin{tabular}{p{.15cm}p{.15cm}p{.6cm}p{.6cm}p{.8cm}p{.15cm}p{.15cm}p{.6cm}p{.6cm}p{.8cm}}\\

\toprule
	 & \multicolumn{4}{c}{Top list} & \multicolumn{5}{c}{TLD} \\
	 & \multicolumn{1}{c}{\textbf{n}}
	 & \multicolumn{1}{c}{\textbf{pre}}
	 & \multicolumn{1}{c}{\textbf{post}}
	 & \multicolumn{1}{c}{\textbf{diff}}
	 &
	 & \multicolumn{1}{c}{\textbf{N}}
	 & \multicolumn{1}{c}{\textbf{pre}}
	 & \multicolumn{1}{c}{\textbf{post}}
	 & \multicolumn{1}{c}{\textbf{diff}}  \\
	\midrule

AT & 455 & 33.0\,\% & 55.2\,\% & 22.2\,\% & .at & 132 & 45.5\,\% & 69.7\,\% & 24.2\,\% \\
BE & 460 & 40.9\,\% & 61.1\,\% & 20.2\,\% & .be & 141 & 59.6\,\% & 78.7\,\% & 19.1\,\% \\
BG & 451 & 37.9\,\% & 60.5\,\% & 22.6\,\% & .bg & 166 & 52.4\,\% & 71.7\,\% & 19.3\,\% \\
CY & 432 & 26.4\,\% & 50.2\,\% & 23.8\,\% & .cy & 58 & 13.8\,\% & 27.6\,\% & 13.8\,\% \\
CZ & 459 & 34.0\,\% & 52.7\,\% & 18.7\,\% & .cz & 251 & 44.6\,\% & 58.2\,\% & 13.5\,\% \\
DK & 447 & 41.2\,\% & 68.9\,\% & 27.7\,\% & .dk & 174 & 72.4\,\% & 87.4\,\% & 14.9\,\% \\
DE & 455 & 26.2\,\% & 49.0\,\% & 22.9\,\% & .de & 172 & 42.4\,\% & 64.5\,\% & 22.1\,\% \\
EE & 441 & 9.5\,\% & 35.8\,\% & 26.3\,\% & .ee & 132 & 14.4\,\% & 35.6\,\% & 21.2\,\% \\
ES & 429 & 41.5\,\% & 64.3\,\% & 22.8\,\% & .es & 86 & 72.1\,\% & 84.9\,\% & 12.8\,\% \\
FI & 462 & 27.5\,\% & 53.9\,\% & 26.4\,\% & .fi & 145 & 37.9\,\% & 55.9\,\% & 17.9\,\% \\
FR & 453 & 49.2\,\% & 66.9\,\% & 17.7\,\% & .fr & 139 & 77.0\,\% & 87.1\,\% & 10.1\,\% \\
GB & 463 & 37.4\,\% & 67.0\,\% & 29.6\,\% & .uk & 108 & 58.3\,\% & 82.4\,\% & 24.1\,\% \\
GR & 443 & 40.0\,\% & 59.8\,\% & 19.9\,\% & .gr & 233 & 56.7\,\% & 69.1\,\% & 12.4\,\% \\
IE & 447 & 21.3\,\% & 64.2\,\% & 43.0\,\% & .ie & 104 & 17.3\,\% & 87.5\,\% & 70.2\,\% \\
IT & 423 & 21.3\,\% & 66.7\,\% & 45.4\,\% & .it & 174 & 30.5\,\% & 90.8\,\% & 60.3\,\% \\
HU & 440 & 46.4\,\% & 62.7\,\% & 16.4\,\% & .hu & 228 & 67.1\,\% & 76.3\,\% & 9.2\,\% \\
HR & 430 & 28.6\,\% & 54.7\,\% & 26.0\,\% & .hr & 141 & 48.9\,\% & 70.9\,\% & 22.0\,\% \\
LV & 434 & 16.8\,\% & 41.9\,\% & 25.1\,\% & .lv & 126 & 38.1\,\% & 61.1\,\% & 23.0\,\% \\
LT & 452 & 27.0\,\% & 47.3\,\% & 20.4\,\% & .lt & 174 & 50.0\,\% & 63.2\,\% & 13.2\,\% \\
LU & 440 & 24.8\,\% & 51.8\,\% & 27.0\,\% & .lu & 61 & 36.1\,\% & 57.4\,\% & 21.3\,\% \\
MT & 446 & 25.8\,\% & 58.1\,\% & 32.3\,\% & .mt & 46 & 21.7\,\% & 43.5\,\% & 21.7\,\% \\
NL & 459 & 37.3\,\% & 54.2\,\% & 17.0\,\% & .nl & 115 & 85.2\,\% & 87.8\,\% & 2.6\,\% \\
PL & 462 & 53.9\,\% & 68.6\,\% & 14.7\,\% & .pl & 256 & 75.4\,\% & 83.2\,\% & 7.8\,\% \\
PT & 430 & 31.4\,\% & 53.7\,\% & 22.3\,\% & .pt & 116 & 52.6\,\% & 65.5\,\% & 12.9\,\% \\
RO & 434 & 30.2\,\% & 53.5\,\% & 23.3\,\% & .ro & 160 & 52.5\,\% & 73.1\,\% & 20.6\,\% \\
SE & 459 & 33.3\,\% & 63.6\,\% & 30.3\,\% & .se & 166 & 50.6\,\% & 78.3\,\% & 27.7\,\% \\
SK & 438 & 42.2\,\% & 56.8\,\% & 14.6\,\% & .sk & 189 & 60.3\,\% & 69.3\,\% & 9.0\,\% \\
SI & 451 & 43.9\,\% & 64.1\,\% & 20.2\,\% & .si & 132 & 75.8\,\% & 77.3\,\% & 1.5\,\% \\
	\midrule
Total & 6357 & 46.1\,\% & 62.1\,\% & 16.0\,\% &   & 4125 & 50.3\,\% & 69.9\,\% & 19.6\,\% \\

\midrule
 &  &  &  &  & .com & 1915 & 28.7\,\% & 50.7\,\% & 22.0\,\% \\
 &  &  &  &  & .net &  248 & 25.4\,\% & 35.5\,\% & 10.1\,\% \\
 &  &  &  &  & .ru  &  148 &  5.4\,\% &  6.7\,\% &  1.3\,\% \\
 &  &  &  &  & .org &  119 & 13.5\,\% & 23.5\,\% & 10.8\,\% \\
 &  &  &  &  & .eu  &   43 & 23.3\,\% & 37.2\,\% & 13.9\,\% \\
 &  &  &  &  & .tr  &   32 &  6.3\,\% &  6.3\,\% &  0.0\,\% \\

	\bottomrule
\end{tabular}
\vspace{-0.5cm}
\end{table}

\subsection{Cookie Banner Libraries}
\label{sec:cb-libraries}

In addition to categorizing the observed cookie notices, we also analyzed commonly encountered third-party cookie libraries in more detail.

During the manual annotation phase of the post-GDPR crawl, we noticed that apart from the increase in usage and complexity of cookie consent notices, the usage of specialized libraries and third parties increased to help websites meet the new legal requirements.
Overall, we identified {31} cookie consent libraries with automated means. We measured their distribution in July 2018 and found that 15.4\,\% of the websites displaying cookie consent notices used one of the identified libraries. Figure~\ref{fig:cookiebanner_distribution_alexa} (b) displays the scores we computed for the different libraries.
\hl{We excluded from our in-depth analysis two libraries not available in English and a WordPress plugin discontinued in November 2018.}

Our results of the analysis of \numCBlibs cookie consent libraries are presented in Table~\ref{table:cb_lib_properties}. We compared the libraries with respect to the following properties:

\emph{Source} identifies whether the code for the consent notice can be hosted by the first party (\emph{self-hosted}) or whether it is retrieved from a \emph{third party}.

\emph{Mechanism} refers to the three distinct mechanisms for consent management. One solution is to have the website asking for consent implement the (un)blocking of cookies according to the user's wishes (\emph{local consent management}). The consent information is stored in a first-party cookie the website can query to react accordingly. \emph{Decentralized consent management} leverages the opt-out APIs provided by third parties, such as online advertisers, to tell them the user's preferences and they are expected to react accordingly. They may remember the user's decision by setting a third-party opt-out cookie. A third option is to use the services of a third party offering \emph{centralized consent management}, who is informed of the user's cookie preferences and triggers the corresponding notifications to participating vendors that would like to set cookies on the user's system. The libraries in our data set that follow this approach have implemented IAB (Interactive Advertising Bureau) Europe's Transparency and Consent Framework. This framework, developed by an industry association, aims to standardize how consent information is presented to the user, collected, and passed down the online advertising supply chain \cite{iab_consent_framework_2018}. IAB-supporting consent notices may display a list of vendors participating in the framework, and the user can select which vendor should be allowed to use their personal data for a variety of purposes. The user selection is encoded in a consent string and transmitted to the participating vendors who committed to comply with the user's selection. \hl{Libraries that do not provide any type of consent management are only capable of displaying a cookie notice.}

Consent notices are presented in one of two ways: \emph{Overlays} block usage of the website until the user clicks one of the banner's buttons. In contrast, \emph{standard banners} are non-modal and thus do not prevent website use while the notice is displayed. Regarding the options the interface may offer to the user, we use the same definitions as in our analysis in Section~\ref{sec:cb-adoption}.

\emph{AutoAccept} refers to mechanisms that automatically assume the user to consent to the use of cookies if they scroll or click a link on the website and react by removing the banner. Some consent libraries offer the website owner to automatically \emph{scan} their site for cookies to assist with sorting them into categories or just display them to provide additional information to the user.

The following two properties are crucial for a library's ability to comply with the user's cookie preferences. The first is the ability to \emph{block cookies}\footnote{For the rest of this section, when we talk about \emph{cookies} in the context of consent, we only refer to cookies that are not considered strictly necessary and thus can only be set with the user's consent.}, \ie, prevent the website from setting cookies if the user has not (yet) consented to their use. If the user changes settings for previously set cookies, the library is expected to \emph{delete cookies}.
\emph{Custom expiration} refers to the site administrator being able to manually set the expiration date of the cookie and thus determine when the consent notice will be shown again. \emph{Geolocation} functionality allows to display the cookie banner only to users from selected areas. 
The \emph{Legal} section lists two properties Article 7 GDPR considers vital for valid consent, the necessity for a data collector to prove that consent was given and the possibility for a user to withdraw consent. If a library allows the user to reconsider and modify their previous consent by displaying a small button or ribbon that opens the consent interface again, we captured this via the \emph{consent change} property. \emph{Consent logging} lets the website owner store information about users' consent decisions for auditing purposes.

Combining the different types of user interfaces with the ability to block and delete cookies allows for the implementation of different types of consent.

\begin{itemize}
\item \emph{Implied Consent} mechanisms assume the user agrees to the use of cookies if they continue to use the website. Implementing this just requires displaying a banner with or without a confirmation button; AutoAccept may also be used. Note that implied consent does not meet the requirements outlined in Article 7 of the GDPR (see \ref{sec:background}).

\item If a site displays a notice

that prevents the user from accessing the site unless the use of cookies is acknowledged, this is referred to as \emph{forced opt-in}. This requires support of the overlay banner type to block access to the website and a confirmation button.
\item An \emph{opt-in} mechanism does not set any non-essential

cookies by default, but users have the opportunity to explicitly allow the use of all the website's cookies. This requires a banner with one (allow) or two (allow / disallow) buttons that blocks cookies by default.
\item In the \emph{opt-out} case, all cookies are set by default, but the user can opt out. This requires the library to display a banner with one (disallow) or two (disallow / allow) buttons and delete cookies that have already been set.
\item More fine-grained types of user selection (slider, checkboxes, individual vendors) just require the library to implement more fine-grained deletion and blocking of cookies. Giving the user more control of which types of cookies to allow and to refuse is in alignment with the GDPR's requirement that consent be given with regard to a specific purpose. It is questionable whether slider-based mechanisms are GDPR-compliant because they force the user to also allow the previous categories in the hierarchy.
\end{itemize}

\begin{table*}
	\centering
	\caption{Properties of cookie consent libraries. \circfill: supports this property, \circempty: does not support this property, B (for ``bug''): functionality exists but did not work, ?: could not be determined, \$: paid version only. * indicates a library we could not install on our test website. \textsuperscript{W}: also available as a WordPress plugin.}
	\label{table:cb_lib_properties}

\begin{threeparttable}
\begin{tabularx}{\linewidth}{lcccccccccccccccccccccc}\\

	\toprule

	& & \multicolumn{2}{c}{Source} & \multicolumn{3}{c}{Mechanism} & \multicolumn{8}{c}{User Interface} & \multicolumn{6}{c}{Technical Details} & \multicolumn{2}{c}{Legal} \\

	\cmidrule(lr){3-4} \cmidrule(lr){5-7} \cmidrule(lr){8-15} \cmidrule(lr){16-21} \cmidrule(lr){22-23}

	 & \textbf{Version} & \rot{\textbf{Self-hosted}} & \rot{\textbf{Third party}} & \rot{\textbf{Local CM}} & \rot{\textbf{Decentralized}} & \rot{\textbf{Centralized}} & \rot{\textbf{Banner}} & \rot{\textbf{Overlay}} & \rot{\textbf{No Option}} & \rot{\textbf{Confirmation}} & \rot{\textbf{Binary}} & \rot{\textbf{Slider}} & \rot{\textbf{Categories}} & \rot{\textbf{Vendors}} & \rot{\textbf{AutoAccept}} & \rot{\textbf{Block Cookies}} & \rot{\textbf{Delete Cookies}} & \rot{\textbf{Cookie Scan}} & \rot{\textbf{Custom Expir.}} & \rot{\textbf{Geolocation}} & \rot{\textbf{Reevaluation}} & \rot{\textbf{Logging}} \\

	\midrule
	\textbf{General Libraries} &&&&&&&&&&&&&&&&&&&&&& \\

	Civic Cookie Control\textsuperscript{W}\tnote{12} & & \circempty & \circfill & \circfill & \circfill & \circempty & \circfill & \circempty & \circempty & \circempty & \circfill & \circempty & \circfill & \circempty & \circempty & \circfill & \circfill & \circempty & \circfill & \$ & \circfill & \circfill \\
   	Clickio Consent Tool*\tnote{13} & & \circempty & \circfill & \circempty & \circempty & \circfill & \circfill & \circfill & \circempty & \circfill & \circfill & \circempty & ? & \circfill & \circfill & \circempty & \circempty & ? & ? & \circfill & \circfill & \circfill \\
   	consentmanager.net\textsuperscript{W}\tnote{14}   & & \circempty & \circfill & \circfill & ? & \circfill & \circfill & \circfill & \circempty & \circfill & \circfill & \circempty & \circfill & \circfill & \circfill & \circfill &  \circempty  & \circfill & \circfill & \circfill & \circfill & \circfill \\
   	cookieBAR\textsuperscript{W}\tnote{15} & 1.7.0 & \circfill & \circfill & \circfill & \circempty & \circempty & \circfill & \circfill & \circempty & \circfill  & \circfill & \circempty & \circempty & \circempty & \circfill & \circempty & \circfill & \circempty & \circfill & \circfill & \circempty & \circempty \\
	Cookiebot\textsuperscript{W}\tnote{16} & & \circempty & \circfill & \circfill & \circempty & \circfill & \circfill & \circfill & \circempty & \circfill & \circfill & \circempty & \circfill & \circempty & \circfill & \circfill & \circfill & \circfill & \circfill & \$ & \circfill & \circfill \\
	Cookie Consent\tnote{17}       & & \circfill & \circfill & \circfill & \circempty  & \circempty & \circfill & \circempty & \circfill & \circfill & \circfill & \circempty & \circempty & \circempty & \circfill & \circfill & \circempty & \circempty & \circfill & \circfill & \circfill & \circempty \\
	\hl{Cookie Information}*\tnote{18} & & \circempty & \circfill & \circfill & \circempty & \circfill & \circfill & \circfill & \circempty & \circfill & \circfill & \circempty & \circfill & \circempty & ? & \circfill & ? & \circfill & ? & ? & \circfill & ? \\
	Cookie Script\tnote{19}*        &  & \$ & \circfill & \circfill & \circempty & \circempty & \circfill & \circfill & \circfill & \circfill & \circfill & \circempty & \circempty & \circempty & \circfill & \circfill & ?  & \circempty & \circfill & \$ & \circfill & \$ \\
	Crownpeak (Evidon)*\tnote{20}  &  & \circempty & \circfill & \circempty & \circfill & \circfill & \circfill & \circfill & \circfill & \circempty  & \circfill & \circempty & \circfill & \circfill & \circfill & \circfill & \circempty & \circfill & \circempty & \circfill & \circempty & \circfill \\
	Didomi*\tnote{21}               & & \circempty & \circfill & \circempty & \circfill & \circfill & \circfill & \circfill & \circempty & \circfill & \circfill  &  \circempty & \circfill & \circfill &  \circempty & \circfill & ?  &  \circempty  & ? & \circfill & \circfill & \circfill  \\
	\hl{jquery.cookieBar}\tnote{22} & & \circfill & \circempty & \circempty & \circempty & \circempty & \circfill & \circempty & \circfill & \circfill & \circempty & \circempty & \circempty & \circempty & \circempty & \circempty & \circempty & \circempty & \circempty & \circempty & \circempty & \circempty \\
	\hl{jQuery EU Cookie Law popups}\tnote{23} & & \circfill & \circempty & \circfill & \circempty & \circempty & \circfill & \circempty & \circfill & \circfill & \circempty & \circempty & \circempty & \circempty & \circfill & \circempty & \circempty & \circempty & \circfill & \circempty & \circempty & \circempty \\
	OneTrust*\tnote{24}            & &  \circempty  & \circfill &  \circfill & \circempty & \circfill & \circfill & ? & \circfill & \circfill & \circempty & \circempty &  \circfill & \circfill & \circfill &  \circfill &  \circfill & \circfill & \circempty & \circfill & \circfill & \circfill \\
	Quantcast Choice\textsuperscript{W}\tnote{25}     &  & \circempty & \circfill & \circempty & \circempty & \circfill & \circfill & \circfill & \circempty & \circfill & \circfill & \circempty & \circfill & \circfill & \circempty & \circempty & \circempty & \circempty & \circfill & \circfill & \circfill &  \circfill \\
	TrustArc (TRUSTe)*\tnote{26}   & & \circempty & \circfill & \circempty  & \circfill & \circfill & \circfill & \circfill & \circfill & \circfill  & \circempty &  \circfill  & \circfill & \circfill & \circfill & \circfill & \circempty  & \circfill & \circempty & \circfill & \circfill & \circfill  \\

	\midrule
	\textbf{WordPress Plugins} &&&&&&&&&&&&&&&&&&&&&& \\

    \hl{Cookie Bar}\tnote{27} & &  \circfill & \circempty & \circempty & \circempty & \circempty & \circfill & \circempty & \circempty & \circfill & \circempty & \circempty & \circempty & \circempty & \circempty & \circempty & \circempty & \circempty & \circempty & \circempty & \circempty & \circempty \\
    Cookie Consent\tnote{28} & 2.3.11 & \circfill & \circempty &  \circempty  & \circempty & \circempty & \circfill & \circempty & \circfill & \circfill & \circempty & \circempty & \circempty & \circempty & \circfill & \circempty & \circempty & \circempty & \circfill & \circfill & \circempty & \circempty\\
    \hl{Cookie Law Bar}\tnote{29} & 1.2.1 & \circfill & \circempty & \circempty & \circempty & \circempty & \circfill & \circempty & \circfill & \circfill & \circempty & \circempty & \circempty & \circempty & \circempty & \circempty & \circempty & \circempty & \circempty & \circempty & \circempty & \circempty \\
    Cookie Notice for GDPR\tnote{30} & 1.2.45 & \circfill & \circempty & \circfill & \circempty & \circempty & \circfill & \circempty &  \circempty & \circfill & \circfill & \circempty & \circempty & \circempty &  \circfill & \circfill & \circempty & \circempty & \circfill & \circempty & \circfill & \circempty \\
    \hl{Custom Cookie Message}\tnote{31} & 2.2.9 & \circfill & \circempty & \circempty & \circempty & \circempty & \circfill & \circempty & \circempty &  \circfill & \circempty & \circempty & \circfill & \circempty & \circempty & \circempty & \circempty & \circempty & \circfill & \circempty & \circempty & \circempty \\
    EU Cookie Law\tnote{32} & 3.0.5 & \circfill & \circempty & \circfill & \circempty & \circempty & \circfill & \circempty & \circempty & \circfill & \circempty & \circempty & \circempty & \circempty & \circfill & \circfill & \circfill & \circempty & \circfill & \circempty & \circfill & \circempty \\
    \hl{GDPR Cookie Compliance}\tnote{33} & 1.2.6 & \circfill & \circempty & \circfill & \circempty & \circempty & \circfill & \$ & \circempty & \circfill & \circempty & \circempty & \circfill & \circempty & \$ & \circfill & \circempty & \circempty & \circempty & \circempty & \circfill & \circempty \\
    GDPR Cookie Consent\tnote{34} & 1.7.1 & \circfill & \circempty & \circfill & \circempty & \circempty & \circfill & \circfill & \circfill & \circfill & \circfill &  \circempty  & \$ & \circempty & \circfill & \circfill & \circfill & ? & \circempty & \$ & \circfill & \$ \\
    \hl{GDPR Tools}\tnote{35} & 1.0.2 & \circfill & \circempty & \circfill & \circempty & \circempty & \circfill & \circempty & \circempty & \circfill & \circfill & \circempty & \circfill & \circfill & \circempty & \$ & ? & \circempty & \$ & \circempty & \circempty & ? \\
    \hl{WF Cookie Consent}\tnote{36}  & 1.1.4 & \circfill & \circempty & \circempty & \circempty & \circempty & \circfill & \circempty & \circempty & \circfill & \circempty & \circempty & \circempty & \circempty & \circempty & \circempty & \circempty & \circempty & \circempty & \circempty & \circempty & \circempty \\

    \midrule
	\hl{\textbf{Drupal Modules}} &&&&&&&&&&&&&&&&&&&&&& \\
	\hl{Cookie Control}\tnote{37} & 1.7-1.6 & \circfill & \circempty & \circfill & \circempty & \circempty & \circfill & \circempty & \circfill & \circempty & \circfill & \circempty & \circempty & \circempty & \circfill & \circfill & \circempty & \circempty & \circempty & B & \circfill & \circempty \\
	\hl{EU Cookie Compliance}\tnote{38} & 7.x-1.25 & \circfill & \circempty & \circfill & \circempty & \circempty & \circfill & \circempty & \circempty & \circfill & \circfill & \circempty & \circempty & \circempty & \circfill & \circfill & ? & \circempty & \circfill & \circfill & \circfill & \circfill \\
	\hl{Simple Cookie Compliance}\tnote{39} & 7.x-1.5 & \circfill & \circempty & \circempty & \circempty & \circempty & \circfill & \circempty & \circempty & \circfill & \circempty & \circempty & \circempty & \circempty & \circempty & \circempty & \circempty & \circempty & \circfill & \circempty & \circempty & \circempty \\

	\bottomrule
\vspace{-0.25cm}
\end{tabularx}
\begin{multicols}{3}
\begin{tablenotes}
\item[12] \url{https://www.civicuk.com/cookie-control}
\item[13] \url{http://gdpr.clickio.com/}
\item[14] \url{https://consentmanager.net}
\item[15] \url{https://cookie-bar.eu}
\item[16] \url{https://cookiebot.com}
\item[17] \url{https://cookieconsent.insites.com}
\item[18] \url{https://cookieinformation.com}
\item[19] \url{https://cookie-script.com}
\item[20] \url{https://evidon.com/solutions/universal-consent/}
\item[21] \url{https://www.didomi.io/en/privacy-center}
\item[22] \url{https://carlwoodhouse.github.io/jquery.cookieBar}
\item[23] \url{https://github.com/wimagguc/jquery-eu-cookie-law-popup}
\item[24] \url{https://onetrust.com/products/cookies}
\item[25] \url{https://quantcast.com/gdpr/consent-management-solution}
\item[26] \url{https://trustarc.com/products/consent-manager}
\item[27] \url{https://wordpress.org/plugins/cookie-bar}
\item[28] \url{https://catapultthemes.com/cookie-consent/}
\item[29] \url{https://wordpress.org/plugins/cookie-law-bar/}
\item[30] \url{https://dfactory.eu/products/cookie-notice/}
\item[31] \url{https://wordpress.org/plugins/custom-cookie-message/}

\item[32] \url{https://wordpress.org/plugins/eu-cookie-law/}
\item[33] \url{https://wordpress.org/plugins/gdpr-cookie-compliance/}
\item[34] \url{https://webtoffee.com/product/gdpr-cookie-consent}

\item[35] \url{https://wordpress.org/plugins/gdpr-tools}
\item[36] \url{https://wordpress.org/plugins/wf-cookie-consent/}
\item[37] An earlier version of Civic Cookie Control for Drupal, \\ \url{https://drupal.org/project/cookiecontrol}
\item[38] \url{https://drupal.org/project/eu_cookie_compliance}
\item[39] \url{https://drupal.org/project/simple_cookie_compliance}

\end{tablenotes}
\end{multicols}
\end{threeparttable}
\vspace{-0.8cm}
\end{table*}
\setcounter{footnote}{39}

Examining the libraries listed in Table~\ref{table:cb_lib_properties}, we made the following observations:

The notion of implied consent is widely supported and easy to implement 
-- adding a banner stating that the website uses cookies just requires adding a JavaScript library to the website or activate a WordPress plugin. The same applies to forced consent. In contrast, types of consent offering the user multiple options require more effort because whether cookies are set and read or not should depend on user consent.

The opt-in scenario can be implemented (a) by overwriting the  \texttt{document.cookie} JavaScript object and add a conditional block that only executes when querying the consent cookie returns that the user has consented. We also found libraries that (b) trigger a JavaScript event when the user has consented, upon which the cookie-setting code is run.
Implementing an opt-out is challenging because it requires the cookie consent library to trigger deletion of the cookies that have already been set. A website can easily delete cookies originating from its own domain -- unless they are \texttt{HttpOnly} or \texttt{Secure} cookies. It cannot delete third-party cookies due to the same-origin policy preventing access to cookies set by another host. Working opt-out mechanisms we found in the (b) scenario use JavaScript events to learn when consent has been revoked for all or selected categories of cookies and then leverage third-party opt-out mechanisms to delete these cookies. Google Analytics, for example, can be triggered to remove its cookies by setting \texttt{window['ga-disable-UA-XXXXXX-Y'] = true}, where \texttt{UA-XXXXXX-Y} references the website ID. This mechanism requires third parties to provide APIs for opt-outs. In case the third party does not,
the user is ideally alerted that their opt-out (partially) failed, as demonstrated by Civic Cookie Control, which displays a warning message that the cookies cannot be deleted automatically and provides a link to the third party's opt-out website. This also poses limitations for cookie settings interfaces: Once a user has agreed to the use of third party cookies, revoking consent is limited to cookies for which deletion can be triggered by the website.

If a library supports consent for different cookie categories, it needs to know which cookies should be considered ``strictly necessary'' such that Art. 5(3) Directive 2002/58/EC applies and consent is not required. If the mapping of cookies into categories is done by the website owner, nothing prevents them from declaring all cookies ``strictly necessary''. We found one notable example on the website of
a major U.S. TV network, where cookies for Google Analytics and Google Ad Serving were categorized as necessary for website operation. One online marketing website used a complex consent solution but had simply declared all cookies necessary, causing the library to merely display a ``no option'' solution.

Fine-grained consent for individual vendors is supported by libraries that implement the IAB framework. The IAB-based consent notices we encountered both provided too much and too little information: By default, the IAB framework's vendor-based cookie selection mechanism displays all of the vendors participating in the framework, not just the ones used by the website.\footnote{\hl{As of December 13, 2018, the IAB supports 460 vendors (\url{https://vendorlist.consensu.org/vendorlist.json}).}}

This renders the fine-grained control offered by the framework unusable. We drew from our dataset a sample of 24 websites with IAB-supporting consent notices (10 Didomi, 7 Clickio, 7 Quantcast) and found that only two sites using Didomi had customized their list of vendors, reducing their number to 21 and 8.

At the same time, the functionality of IAB-based consent notices is limited to IAB vendors, unless the library also supports other vendors as in Didomi's consent mechanism, which has integrated additional vendors including Google and Facebook. As we observed during the manual annotation of consent notices, IAB banners tend to display a standard text that does not inform users that the website may also use other third parties in addition to listed IAB vendors and that those other parties are not bound by the user's consent decision made in the IAB-based tool.

Our analysis shows that implementing GDPR consent requirements in practice with existing libraries is a challenge. The GDPR's requirements for informed consent include an affirmative action by the user upon having been provided with sufficient information about the purposes of cookie use. This is at odds with usability as studies have shown the ineffectiveness of previous choices mechanisms~\cite{leon2012johnny}.

The options to implement meaningful choices for the user, including the ability to withdraw consent, are limited by technical restrictions, such as the same-origin policy, a core principle of web security, and the business interests of third parties, not all of which are interested in providing an opt-out API.
Under the GDPR, consent has to be given for specific purposes of data processing, which raises the question who defines the purpose of the use of a certain cookie. If left to the developers or site owners, it is prone to abuse of the ``strictly necessary'' category to circumvent the consent requirement in Directive 2002/58/EC.

\section{Discussion and Future Work}

\fixme{Our results show that at the time the GDPR came into force websites made changes that can be considered improvements for web privacy}, but the goal of harmonization is not yet met. We discuss resulting challenges and opportunities for researchers, policymakers, and companies. We also discuss some limitations of our study.

\subsection{Impact of the GDPR}

Our analysis focuses on the 28 EU member states, but the GDPR also impacts websites from other countries -- first because some non-EU countries have decided to adopt similar rules (\eg, Norway, Switzerland, Iceland and Liechtenstein~\cite{efta})

and second, because websites that offer services in the EU have to comply with the GDPR. For example, according to Alexa, 53\% of the U.S. top 500 websites and 48\% of the most visited Russian sites also appear in at least one EU state's top 500 list.
A positive finding of our study is that even though the majority of websites already had privacy policies, the prevalence of privacy policies increased even further. Our results suggest that the harmonization of data protection rules could eventually lead to consistent privacy policy adoption rates across Europe. We also see the increased mention of GDPR-specific terms across all countries as a sign for the GDPR's impact and a step towards harmonization. However, despite this trend, actions taken to comply with GDPR vary greatly, especially regarding consent and cookies.

\subsection{Need for More Detailed and Practical GDPR Guidance}

Although the GDPR makes it clear that websites require a privacy policy, details about what is permissible or required remain unclear. Especially with respect to cookie consent notifications, the observed variance in implementation indicates the need for clearer guidelines for service providers. Such guidance should, for example, clarify what types of cookies can be set on what legal grounds. This requires determinations on questions such as whether website operators can claim a ``legitimate interest'' in web analytics or if user tracking requires explicit consent.

There is hope that a future ePrivacy Regulation may provide some clarity regarding these issues, but at the time of writing it is unclear when and it what form it may be adopted.
Our results also show that some countries lag behind in the adoption of privacy policies. To improve the situation, data protection authorities could support companies by providing effective means for cookie handling, consent mechanisms, and privacy statements.

\subsection{False Sense of Compliance}
Some of this uncertainty about how to interpret the GDPR may result in a false sense of compliance.
Although the majority of websites in our dataset now have an up-to-date privacy policy, 15.5\,\% still do not have one and 14.9\,\% have not updated it in the last years. While the prevalence of privacy policies in the finance or shopping sector is close to 100\,\% and we do not expect semi-legal services in the streaming sector to be compliant, a number of websites in news, business, or education are likely not compliant with GDPR.
Companies should also be aware that the widely used cookie banners that only inform users are not sufficient to obtain users' consent. As the Article 29 working group stated, \textit{``merely proceeding with a service cannot be regarded as an active indication of choice''}~\cite{article_29_consent}. After all, companies violating GDPR risk fines of up to 4\,\% of their worldwide annual turnover.

\subsection{Opportunities for Web Privacy and Security Research}
The presence of a privacy policy does not mean that a service is compliant with privacy law. More research is needed to study whether a privacy policy's content actually meets legal requirements. So far, research on web privacy has largely been focused on English-language privacy policies and web users. Our study shows differences among countries and suggests that rather tiny language communities would benefit from a more multi-lingual research approach.  Thus, the GDPR creates an interesting environment for privacy and security research not just to study its implementation but also to evaluate new ideas on how to improve security and privacy online. GDPR requires service providers to use ``state-of-the-art technology'' and our results indicate that the GDPR has already fostered increased adoption of HTTPS and cookie consent mechanisms.
The increased prevalence of privacy policies as natural language descriptions of data practices, with more technical approaches like Do Not Track and P3P failing at the same time, increases the need for research that closes the gap between legal and technical privacy means. Research could help to raise minimum security standards by creating new, easy to adopt security mechanisms and improve usability with browser-based implementations of consent mechanisms.
\hl{To foster research in this area, the tools and data sets used for this study are publicly available in a GitHub repository.\footnote{\url{https://github.com/RUB-SysSec/we-value-your-privacy}.}}

\section{Related Work}

Privacy policies have been studied extensively as they constitute one of the primary means of transparency. While few have studied longitudinally the prevalence of privacy policies, prior work has analyzed how they are perceived by users, what they disclose, and how they present information to users.

\subsection{Adoption of Privacy Policies}
The U.S. Federal Trade Commission
first evaluated the use of privacy policies in 1998 and found that only 14\,\% of 674 websites studied had a privacy policy~\cite{ftc_1998}. Numbers had increased when Liu \& Arnett in 2002 received a privacy policy from 64\,\% of companies~\cite{liu_raising_2002}.
In 2017, Nokhbeh \& Barber~\cite{nokhbeh_zaeem_study_2017} found that of the 600 biggest companies by stock value 70\,\% had a privacy policy. Both studies were based on stock exchange listings, not popularity online. Both found huge differences between industry sectors, with the technology sector among the ones with higher privacy policy adoption rates of around 80\,\%.
Story et al. examined one million Android apps in the U.S. Google Play Store and found that the percentage featuring privacy policies had increased from 41.7\,\% in September 2017 to 51.8\,\% in mid-May 2018~\cite{story_apps_2018}.

\subsection{Usefulness of privacy policies}
Researchers have also studied privacy policies' content and how users deal with these increasingly complex documents. McDonald and Cranor~\cite{mcdonald_cost_2008} concluded that a typical web user would have to spend 244 hours annually if they wanted to read every privacy policy of the websites they visit; it would further require a college degree to actually understand them \cite{proctor_examining_2008}. Obar et al. recently confirmed that few people open privacy policies or terms of service they agree to when registering for a service, and over 90\,\% miss important details \cite{obar_biggest_2018}.
Still, reading privacy policies can help consumers build trust in companies~\cite{ermakova_privacy_2014}, although recently Turow et al. \cite{turow_persistent_2018} published a meta-study and showed that the pure existence of a privacy policy seems to be sufficient to achieve this goal, due to misconceptions of companies' data practices.

Such misconceptions are even higher for younger adults.

\subsection{Analysis of Privacy Policies}
Based on the results about the usefulness of privacy policies, researchers have started to support users and make privacy policies easier to comprehend or completely automate their assessment. To support machine learning approaches, Wilson et al.~\cite{wilson_creation_2016} created a corpus of 115 privacy policies of U.S. companies, which was extensively annotated by law students to identify described data practices.

Harkous et al.~\cite{harkous_polisis:_2018} used the same corpus to train a deep learning system that allows querying privacy policies with natural language questions. Gluck et al.~\cite{gluck_how_2016} evaluated how the length of privacy notices affects awareness of certain practices and concluded that (automatically) shortening privacy policies has potential, but important aspects may get lost if not done carefully. Leveraging the design space for privacy notices and controls may help create concise and actionable notices with integrated choice~\cite{schaub2017,schaub2015}.
Other researchers aim to extract information from privacy policies. Libert \cite{libert_2018} analyzed English-language privacy policies to automatically check whether they disclose the names of companies doing third-party tracking on websites. Sathyendra et al. \cite{sathyendra_automatic_2016} evaluated how the options users have, especially about opting out, can automatically be identified in privacy policies. Tesfay et al. \cite{tesfay_privacyguide_2018} collected privacy policies from the top 50 websites in Europe as identified by the Alexa ranking and developed a tool to summarize them and visualize the results inspired by GDPR criteria.

All these approaches currently focus on English-language documents as English

dominates the Web. Few researchers have evaluated other or multiple languages. Fukushima et al. \cite{fukushima_2018} evaluated machine learning approaches on a set of annotated Japanese privacy policies and found that automatic classifiers struggle with identifying important sections due to redundancy in the language. Cha \cite{cha_information_2011} compared privacy policies of Korean and U.S. websites based on the rules set by the EU privacy directive and found Korean websites to provide stronger privacy policies, but also to request more data from their users. To the best of our knowledge, no prior studies have evaluated and compared privacy policies from numerous countries, let alone all EU member states.

\subsection{Cookie Consent Notices}
Taking into account that cookie consent notices are not supposed to be necessary (see Section~\ref{sec:background}), research on them is scarce.
In February 2015, the Article 29 Working Party conducted a ``Cookie Sweep'' to determine the effects of Directive 2009/136/EC's requirements~\cite{article_29_wp_cookie_sweep_2015}. In eight EU member states, 437 sites were manually inspected for information they provided about cookies, including the type and position of the interface used. At that time, 116 (26\,\%) of the analyzed sites did not provide any information about cookie use; for another 39\,\% the information was deemed not sufficiently visible. Of the remaining 404 sites, 50.5\,\% (204) sites were found to \textit{``request [...] consent from the user to store cookies''} while 49.5\,\% (200) simply stated that cookies were being used. 16\,\% (49 sites) offered the user to accept or decline certain types of cookies. The study did not investigate whether the banners asking for consent implemented a proper opt-in mechanism.
More recently, Kulyk et al. \cite{oksana_kulyk_this_2018} collected cookie consent notices from the top 50 German websites in the Alexa ranking to investigate how users perceive and react to different types of banners. They identified five distinct groups of notices based on the amount of information they provide about 
cookie use but did not analyze users' options for interacting with the banner.

\section{Conclusion}

Our analysis of the top 500 websites in each of the EU member states, involving the analysis of privacy policies in 24 languages, \fixme{indicate positive effects on web privacy taking place around the GDPR enforcement date}. While most websites already had privacy policies, a large majority made adjustments. Most notable is the rise of cookie consent banners, which now greet European web users on more than half of all websites. While seemingly positive, the increase in transparency may lead to a false sense of privacy and security for users. Few websites offer their users actual choice regarding cookie-based tracking. Moreover, most of the analyzed cookie consent libraries do not meet GDPR requirements.

Browser manufacturers and the industry so far have not been able to agree on technical privacy standards, such as Do Not Track. This puts an additional burden on users, who are presented with an increasing number of privacy notifications that may fulfill the law's transparency requirements but are unlikely to actually help web users make more informed decisions regarding their privacy. In addition, regulators need to provide clear guidelines in what cookies a service can claim ``legitimate interests'' and which should require actual consent.

\endgroup


\section*{Acknowledgments}
The authors would like to thank Yana Koval for her help with manual website annotation and all native speakers who helped us verify the word lists.
This research was partially funded by the MKW-NRW Research Training Groups SecHuman and NERD.NRW, and the National Science Foundation under grant agreement CNS-1330596.






%


\balance

\bibliographystyle{IEEEtranS}
\bibliography{bibliography}




\balance
\onecolumn
\section{Appendix}
\label{sec:appendix}
\begin{table*}[ht]

	\caption{Countries and codes}
    \begin{tabular}{p{2cm}p{.5cm}p{.5cm}p{1cm}p{6cm}p{6cm}}\\

	\textbf{Country} &\textbf{Code} &\textbf{TLD} &\textbf{Lang} & \textbf{Words identifying links to privacy policies} & \textbf{GDPR} \\

	\toprule

Austria & AT & .at & DE & datenschutz, datenrichtlinie & see DE\\
\midrule
Belgium & BE & .be & NL,FR,DE & see FR/NL/DE & see FR/NL/DE \\
\midrule
Bulgaria & BG & .bg & BG & \selectlanguage{bulgarian}поверителност, политика за данни, политика за бисквитки
& \selectlanguage{bulgarian}Закона за електронната търговия , Общ регламент относно защитата на данните\\
\midrule
Cyprus & CY & .cy & EL, TR & gizlilik, veri ilkesi, see EL

\\
\midrule
Czech Republic & CZ & .cz & CS & soukrom\'{i}, z\'{a}sady pou\v{z}\'{i}v\'{a}n\'{i} dat, ochrana soukrom\'{i}, podm\'{i}nky, ochrana dat, ochrana osobn\'{i}ch \'{u}daj\r{u} &  obecn\'{e} na\v{r}ízen\'{i} o ochran\v{e} osobn\'{i}ch \'{u}daj\o{o}

\\
\midrule
\selectlanguage{english}
Germany & DE & .de & DE & datenschutz, privatsph\"{a}re, datenschutzbestimmungen, datenschutzrichtlinie & Datenschutzgrundverordnung \\
\midrule
Denmark & DK & .dk & DA & beskyttelse af personlige oplysninger, datapolitik, cookiepolitik, privatlivspolitik, personoplysninger, regler om fortrolighed, personlige data & generel forordning om databeskyttelse \\
\midrule

Estonia & ET & .ee & ET & privaatsus,data policy, isikuandmete, isikuandmete t\"{o}\"{o}tlemise, k\"{u}psised, konfidentsiaalsuse, andmekaitsetingimused & isikuandmete kaitse \"{u}ldm\"{a}\"{a}rus \\
\midrule

Spain & ES & .es & ES & privacidad, pol\'{i}tica de datos, protecció de dades, aviso legal & Reglamento general de protección de datos

\\
\midrule
Finland & FI & .fi & FI & yksityisyys, tietok\"{a}yt\"{a}ntö, tietosuojak\"{a}yt\"{a}ntö, yksityisyyden suoja, tietosuojaseloste, rekisteriseloste, tietosuoja, yksityisyydensuoja & yleinen tietosuoja-asetus \\

\midrule
France & FR & .fr & FR & confidentialit\'{e}, politique d’utilisation des donn\'{e}es,  mentions l\'{e}gales, cgu, cookies, vie priv\'{e}e, donnees personelles, mentions l\'{e}gales & r`{e}glement g\'{e}n\'{e}ral sur la protection des donn\'{e}es\\
\midrule

Greece & GR & .gr & EL & \selectlanguage{greek}απόρρητο, όροι και γνωστοποιήσεις, προσωπικά δεδομένα, πολιτική απορρήτου &  \selectlanguage{greek}Γενικός Κανονισμός για την Προστασία Δεδομένων \\

\midrule
Croatia & HR & .hr & HR & privatnost, privatnosti, pravila o upotrebi podataka, za\v{s}tita podataka, kolači\'{c}i & Op\'{c}a uredba o za\v{z}titi podataka
 \\
 \midrule
Hungary & HU & .hu & HU & adatv\'{e}delem, adatkezel\'{e}si, adatv\'{e}delmi, szem\'{e}lyes adatok v\'{e}delme & \'{a}ltal\'{a}nos adatv\'{e}delmi rendelet
 \\
 \midrule
Ireland  & IE & .ie &  GA,EN & see EN & An Rialachán Ginearálta maidir le Cosaint Sonraí \\
\midrule
Italy & IT & .it & IT & normativa sui dati &  regolamento generale sulla protezione dei dati \\
\midrule

Lithuania & LT & .lt & LT & privatumas, slapukai, privatumo & Bendrasis duomen\k{u} apsaugos reglamentas \\
\midrule
Luxembourg & LU & .lu & DE/FR & see DE, FR &  see DE, FR \\
\midrule

Latvia & LV & .lv & LV & priv\={a}tums, priv\={a}tuma, sīkdatņu, sīkdatne & Visp\={a}r\={i}g\={a} datu aizsardz\={i}bas regula \\
\midrule
Malta & MT & .mt & MT & privatezza & Regolament \.{G}enerali dwar il-Protezzjoni tad-Data\\
\midrule
Netherlands & NL & .nl & NL & gegevensbeleid, privacybeleid & algemene verordening gegevensbescherming \\
\midrule
Poland & PL & .pl & PL & prywatno\'{s}\'{c}, zasady dotyczące danych, prywatno\'{s}ci & og\'{o}lne rozporz\k{a}dzenie o ochronie danych \\
\midrule
Portugal & PT & .pt &  PT & privacidade, pol\'{i}tica de dados & Regulamento Geral sobre a Proteção de Dados \\
\midrule
Romania & RO & .ro &  RO & confidențialitate, politica de utilizare, cookie-uri, confidentialitate, cookie-urilor, protecţia datelor & Regulamentul general privind protecția datelor \\

\midrule
Slovakia & SK & .sk & SK & ochrana s\'{u}kromia,z\'{a}sady vyu\v{z}\'{i}vania \'{u}dajov, ochrana \'{u}dajov, ochrana osobn\'{y}ch \'{u}dajov, s\'{u}kromie, pi\v{s}kotki, z\'{a}sady ochrany osobn\'{y}ch & všeobecné nariadenie o ochrane údajov\\

\midrule
Slovenia & SI  & .si & SL & zasebnost, pi\v{s}kotkih, varstvo podatkov & Splošna uredba o varstvu podatkov \\

\midrule
Sweden & SE & .se & SV & sekretess, datapolicy, personuppgifter, webbplatsen, integritetspolicy & allmän dataskyddsförordning \\
\midrule

United Kingdom & UK & .uk & EN & privacy, privacy policy & General Data Protection Regulation \\
	\bottomrule
	\label{table:ppwords}
\end{tabular}
\end{table*}

\begin{table*}[ht]
\label{table:wording2}
	\caption{List of GDPR Phrases I}
    \begin{tabular}{p{2.5cm}p{2.5cm}p{2.5cm}p{2.5cm}p{2.5cm}p{2.5cm}}\\
    \toprule
	BG & CS & DE & EN & EL & ES\\

	\toprule

 \selectlanguage{bulgarian}\vspace{-0.1cm}администратор & spr\'{a}vca & Verantwortliche & controller & \selectlanguage{greek}\vspace{-0.2cm}υπεύθυνος επεξεργασίας & responsable\\
 \midrule
 \selectlanguage{bulgarian}\vspace{-0.1cm}длъжностното лице по защита на данните & pov\v{e}\v{r}enec pro ochranu osobn\'{i}ch \'{u}daj\r{u} & Datenschutzbeauftragte & data protection officer & \selectlanguage{greek}\vspace{-0.2cm}υπεύθυνος προστασίας δεδομένων & delegado de protecci\'{o}n de datos\\
 \midrule
 \selectlanguage{bulgarian}\vspace{-0.1cm}цел & \'{u}\v{c}el & Zweck & purposes & \selectlanguage{greek}\vspace{-0.2cm}σκοπός & fin\\
 \midrule
 \selectlanguage{bulgarian}\vspace{-0.1cm}правното основание & pr\'{a}vn\'{i} z\'{a}klad & Rechtsgrundlage & legal basis & \selectlanguage{greek}\vspace{-0.2cm}νομική βάση  & base jur\'{i}dica\\
 \midrule
 \selectlanguage{bulgarian}\vspace{-0.1cm}обработване & zpracov\'{a}n\'{i} & Verarbeitung & processing & \selectlanguage{greek}\vspace{-0.2cm}επεξεργασία & tratamiento\\
 \midrule
 \selectlanguage{bulgarian}\vspace{-0.1cm}законните интереси & opr\'{a}vn\v{e}n\'{e} z\'{a}jmy & berechtigte Interessen & legitimate interests & \selectlanguage{greek}\vspace{-0.2cm}έννομα συμφέροντα & intereses leg\'{i}timos\\
  \midrule
 \selectlanguage{bulgarian}\vspace{-0.1cm}получателите & p\v{r}\'{i}jemce & Empf\"{a}nger & recipients &\selectlanguage{greek}\vspace{-0.2cm}αποδέκτης & destinatarios\\
 \midrule
 \selectlanguage{bulgarian}\vspace{-0.1cm}трета държава & t\v{r}et\'{i} zem\v{e} & Drittland & third country & \selectlanguage{greek}\vspace{-0.2cm}τρίτη χώρα & tercer pa\'{i}s\\
 \midrule
 \selectlanguage{bulgarian}\vspace{-0.1cm}срок & doba & Dauer & period & \selectlanguage{greek}\vspace{-0.2cm}χρονικό διάστημα & plazo\\
 \midrule
 \selectlanguage{bulgarian}\vspace{-0.1cm}информация & p\v{r}\'{i}stup & Auskunft & access & \selectlanguage{greek}\vspace{-0.2cm}πρόσβαση & acceso\\
 \midrule
 \selectlanguage{bulgarian}\vspace{-0.1cm}коригиране & oprava & Berichtigung & rectification & \selectlanguage{greek}\vspace{-0.2cm}διόρθωση & rectificaci\'{o}n\\
 \midrule
 \selectlanguage{bulgarian}\vspace{-0.1cm}изтриване & výmaz & L\"{o}schung & erasure & \selectlanguage{greek}\vspace{-0.2cm}διαγραφή  & supresi\'{o}n\\
 \midrule
 \selectlanguage{bulgarian}\vspace{-0.1cm}ограничаване & omezen\'{i}  & Einschr\"{a}nkung & restriction & \selectlanguage{greek}περιορισμός & limitaci\'{o}n\\
 \midrule
 \selectlanguage{bulgarian}\vspace{-0.1cm}възражение & pr\'{a}vo vzn\'{e}st n\'{a}mitku & Widerspruchsrecht & object & \selectlanguage{greek}\vspace{-0.2cm}αντίταξης  & oponerse\\
 \midrule
 \selectlanguage{bulgarian}\vspace{-0.1cm}преносимост на данните & p\v{r}enositelnost \'{u}daj\r{u} & Daten\"{u}bertragbarkeit & data portability & \selectlanguage{greek}\vspace{-0.2cm}φορητότητα δεδομένων & portabilidad de los datos\\
 \midrule
 \selectlanguage{bulgarian}\vspace{-0.1cm}оттегляне на съгласието & odvolat souhlas & Einwilligung widerrufen & withdraw consent & \selectlanguage{greek}\vspace{-0.2cm}ανακαλώ τη συγκατάθεσή & retirar el consentimiento\\
 \midrule
 \selectlanguage{bulgarian}\vspace{-0.1cm}жалба & st\'{i}\v{z}nost & Beschwerde & complaint & \selectlanguage{greek}\vspace{-0.2cm}καταγγελία & reclamaci\'{o}n\\
 \midrule
 \selectlanguage{bulgarian}\vspace{-0.1cm}надзорен орган & dozorový \'{u}\v{r}ad & Aufsichtsbeh\"{o}rde & supervisory authority & \selectlanguage{greek}\vspace{-0.2cm}εποπτική αρχή  & autoridades de control \\
 \midrule
 \selectlanguage{bulgarian}\vspace{-0.1cm}договор & smlouva & Vertrag & contract & \selectlanguage{greek}\vspace{-0.2cm}σύμβαση & contrato\\
 \midrule
 \selectlanguage{bulgarian}\vspace{-0.1cm}задължително изискване & z\'{a}konný po\v{z}adavek & gesetzlich vorgeschrieben & statutory requirement & \selectlanguage{greek}\vspace{-0.2cm}νομική υποχρέωση & requisito legal\\
 \midrule
 \selectlanguage{bulgarian}\vspace{-0.1cm}договорно изискване & smluvn\'{i} po\v{z}adavek & vertraglich vorgeschrieben & contractual requirement & \selectlanguage{greek}\vspace{-0.2cm}συμβατική υποχρέωση & requisito contractual\\
 \midrule
 \selectlanguage{bulgarian}\vspace{-0.1cm}последствия & d\r{u}sledek & Folgen & consequences & \selectlanguage{greek}\vspace{-0.2cm}συνέπεια & consecuencias \\
 \midrule
 \selectlanguage{bulgarian}\vspace{-0.1cm}автоматизирано вземане на решения & automatizovan\'{e} rozhodov\'{a}n\'{i} & automatisierte Entscheidungsfindung & automated decision-making & \selectlanguage{greek}\vspace{-0.2cm}αυτοματοποιημένη λήψη αποφάσεων & decisiones automatizadas\\
 \midrule
 \selectlanguage{bulgarian}\vspace{-0.1cm}профилирането & profilov\'{a}n\'{i} & Profiling & profiling & \selectlanguage{greek}\vspace{-0.2cm}κατάρτιση προφίλ & elaboraci\'{o}n de perfiles\\
 \midrule
 \selectlanguage{bulgarian}\vspace{-0.1cm}по-нататъшно обработване & dal\v{s}\'{i} zpracov\'{a}n\'{i} & Weiterverarbeitung & further processing & \selectlanguage{greek}\vspace{-0.2cm}περαιτέρω επεξεργασία & tratamiento ulterior\\
 \midrule
 \selectlanguage{bulgarian}\vspace{-0.1cm}съгласие & souhlas & Einwilligung & consent & \selectlanguage{greek}\vspace{-0.2cm}συγκατάθεση & consentimiento\\
 \midrule
 \selectlanguage{bulgarian}\vspace{-0.1cm}изпълнение на договор & spln\v{e}n\'{i} smlouvy & Erf\"{u}llung eines Vertrags & performance of a contract & \selectlanguage{greek}\vspace{-0.2cm}εκτέλεση σύμβασης & ejecutar un contrato\\
 \midrule
 \selectlanguage{bulgarian}\vspace{-0.1cm}законово задължение & pr\'{a}vna povinnost & rechtliche Verpflichtung & legal obligation & \selectlanguage{greek}\vspace{-0.2cm}έννομη υποχρέωση & obligaci\'{o}n legal\\
 \midrule
 \selectlanguage{bulgarian}\vspace{-0.1cm}жизненоважни интереси & \v{z}ivotn\v{e} d\r{u}le\v{z}itý z\'{a}jem  & lebenswichtiges Interesse & vital interest & \selectlanguage{greek}\vspace{-0.2cm}ζωτικό συμφέρον & inter\'{e}s vital\\
 \midrule
 \selectlanguage{bulgarian}обществен интерес & ve\v{r}ejný z\'{a}jem & \"{o}ffentliches Interesse & public interest & \selectlanguage{greek}\vspace{-0.2cm}δημόσιο συμφέρον & inter\'{e}s p\'{u}blico\\
 \midrule
 \selectlanguage{bulgarian}\vspace{-0.1cm}официално правомощие & ve\v{r}ejn\'{a} moc  & \"{o}ffentliche Gewalt & official authority & \selectlanguage{greek}\vspace{-0.2cm}δημόσια εξουσία & poder p\'{u}blico\\
 \midrule
 \selectlanguage{bulgarian}\vspace{-0.1cm}публичен орган & org\'{a}n ve\v{r}ejn\'{e} moci & Beh\"{o}rde & public authority & \selectlanguage{greek}\vspace{-0.2cm}δημόσια αρχή & autoridad\\
 	\bottomrule
\end{tabular}
\end{table*}

\begin{table*}[ht]
\label{table:wording3}
	\caption{List of GDPR Phrases II}
    \begin{tabular}{p{1.6cm}p{1.6cm}p{1.6cm}p{1.6cm}p{1.6cm}p{1.6cm}p{1.6cm}p{1.6cm}p{1.6cm}}\\
    \toprule
ET & FI & FR & GA & HR & HU & IT & LV & LT\\

	\toprule

 vastutav t\"{o}\"{o}tleja  & rekisterinpit\"{a}j\"{a} & responsable du traitement & rialaitheoir & voditelj obrade & adatkezel\H{o} & titolare del trattamento & pārzinis & duomenų valdytojas\\
  \midrule
 andmekaitseametnik & tietosuojavastaava & d\'{e}l\'{e}gu\'{e} \`{a} la protection des donn\'{e}es & oifigeach cosanta sonra\'{i} & slu\v{z}benik za za\v{s}titu podataka & adatv\'{e}delmi tisztvisel\H{o}  & responsabile della protezione dei dati & datu aizsardz\~{i}bas speciālistu & duomenų apsaugos pareigūnas\\
  \midrule
  eesm\"{a}rk & tarkoitus & finalit\'{e}s & cr\'{i}ocha & svrh & c\'{e}l & finalit\`{a} & nolūks & tikslas\\
  \midrule
  \~{o}iguslik alus & oikeusperuste & base juridique & bun\'{u}s dl\'{i} & pravna osnova & jogalap & base giuridica & juridiskais pamats & teisinį pagrind\k{a}\\
  \midrule
  t\"{o}\"{o}tlemine & k\"{a}sittely & traitement & pr\'{o}ise\'{a}il & obrada & adatkezel\'{e}s & trattamento & apstrāde & duomenų tvarkymas\\
  \midrule
  \~{o}igustatud huvi & oikeutetut edut & int\'{e}r\u{e}ts l\'{e}gitimes & leasanna dlisteanacha & legitimne interese & jogos \'{e}rdek & legittimo interesse & le\.{g}it\~{i}mās intereses & teis\.{e}tas interesas\\
  \midrule
  vastuv\~{o}tja & vastaanottajat & destinataires & faighteoir\'{i} & primatelje & c\'{i}mzettek & destinatario & saņēmējs & duomenų gav\.{e}jas\\
  \midrule
  kolmas riik & kolmas maa & pays tiers  & tr\'{i}\'{u} t\'{i}r & treća zemlja & harmadik orsz\'{a}g & paese terzo & tre\v{s}ā valsts & tre\v{c}ioji valstyb\.{e}\\
  \midrule
  ajavahemik & s\"{a}ilytysaika & dur\'{e}e & tr\'{e}imhse & razdoblje & id\H{o}tartalom & periodo & laikposms & laikotarpis\\
  \midrule
  juurdep\"{a}\"{a}s & p\"{a}\"{a}sy & acc\`{e}s & rochtain & pristup & hozz\'{a}f\'{e}r\'{e}s & accesso & piekļuve & prieiga\\
  \midrule
  parandamine & oikaisu & rectification & ceart\'{u} & ispravak & helyesb\'{i}t\'{e}s & rettifica & labo\v{s}ana & i\v{s}taisyti\\
  \midrule
  kustutamine & poistaminen & effacement & scriosadh & brisanje & t\"{o}rl\'{e}s & cancellazione & dzē\v{s}anu & i\v{s}trinti\\
  \midrule
  piiramine & rajoitus & limitation & srian & ograni\v{c}avanje & korl\'{a}toz\'{a}s & limitazione & ierobe\v{z}o\v{s}anu & apriboti\\
  \midrule
  vastuv\"{a}ide & vastustaa & s'opposer & ag\'{o}id a dh\'{e}anamh & ulaganje prigovora & tiltakozni & opporsi & iebilst  & nesutikti \\
  \midrule
  andmete \"{u}lekandmine & tietojen siirto & portabilit\'{e} des donn\'{e}es & iniomparthacht sonra\'{i} & prenosivost podataka & az adat hordozhat\'{o}s\'{a}g & portabilit\`{a} dei dati & datu pārnesam\~{i}ba & duomenų perkeliamumas\\
  \midrule
  n\~{o}usolek tagasi v\~{o}tta & peruuttaa suostumus & retirer consentement & toili\'{u} a tharraingt siar & povu\v{c}iti privolu & hozz\'{a}j\'{a}rul\'{a}s visszavon\'{a}sa & revocare il consenso & atsaukt piekri\v{s}anu & at\v{s}aukti sutikim\k{a}\\
  \midrule
  kaebus  & valitus & r\'{e}clamation & gear\'{a}n & prigovor & panasz  & reclamo & sūdz\~{i}ba & skund\k{a}s\\
  \midrule
  j\"{a}relevalveasutus & valvontaviranomainen & autorit\'{e} de contrôle & \'{u}dar\'{a}s maoirseachta & nadzorno tijelo & fel\"{u}gyeleti hat\'{o}s\'{a}gk\'{e}nt & autorit\`{a} di controllo  & uzraudz\~{i}bas iestāde & prie\v{z}iūros institucija\\
  \midrule
  leping & sopimus & contrat  & conradh & ugovor & szerz\H{o}d\'{e}s  & contratto & l\~{i}gums & sutartis\\
  \midrule
  \~{o}igusaktist tulenev kohustus  & lakis\"{a}\"{a}teinen vaatimus  & caract\`{e}re r\'{e}glementaire & ceanglas reachtach & zakonska obveza & jogszab\'{a}lyos k\"{o}telezetts\'{e}g & obbligo legale & noteikta ar likumu  & teis\.{e}s reikalavimas \\
  \midrule
  lepingust tulenev kohustus & sopimuksellinen vaatimus & caract\`{e}re contractuel & ceanglas conarthach & ugovorna obveza & szerz\H{o}d\'{e}ses k\"{o}telezetts\'{e}g & obbligo contrattuale & noteikta ar l\~{i}gumu & sutartyje numatytas reikalavimas\\
  \midrule
  tagaj\"{a}rg & seuraukset & cons\'{e}quences & hiarmhairt\'{i} & posljedice & k\"{o}vetkezm\'{e}nyek & conseguenza & sekas & pasekm\.{e}s\\
  \midrule
  automatiseeritud otsuste tegemine & automaattinen p\"{a}\"{a}t\"{o}ksenteko & prise de d\'{e}cision automatis\'{e}e & chinnteoireacht uathoibrithe & automatizirano dono\v{s}enje odluka & automatiz\'{a}lt d\"{o}nt\'{e}shoz\'{a}s & processo decisionale automatizzato & automatizēta lēmumu pieņem\v{s}ana & automatizuotas sprendimų pri\.{e}mimas\\
  \midrule
  profiilianal\"{u}\"{u}s & profilointi & profilage & pr\'{o}if\'{i}li\'{u} & izrada profila & profilalkot\'{a}s & profilazione & profilē\v{s}ana & profiliavimas\\
  \midrule
  edasine t\"{o}\"{o}tlemine & jatkok\"{a}sittely & traitement ult\'{e}rieur & phr\'{o}ise\'{a}il tuilleadh & dodatno obrađivati & tov\'{a}bbi adatkezel\'{e}s & ulteriore trattamento & turpmāk apstrādāt & tolesnis tvarkymas\\
  \midrule
  n\~{o}usolek & suostumus & consentir & toili\'{u} & privola & hozz\'{a}j\'{a}rul\'{a}s  & consenso & piekri\v{s}anu & sutikim\k{a}\\
  \midrule
  lepingu t\"{a}itmine & sopimuksen t\"{a}ytt\"{a}minen & ex\'{e}cution d'un contrat & comhl\'{i}onadh conartha & izvr\v{s}avanje ugovora & szerz\H{o}d\'{e}s teljes\'{i}t\'{e}s & esecuzione di un contratto & l\~{i}guma izpilde & sutarties vykdymas\\
  \midrule
  juriidiline kohustus & lakis\"{a}\"{a}teinen velvoite & obligation l\'{e}gale & oibleag\'{a}id dhl\'{i}thi\'{u}il  & pravna obveza & jogi k\"{o}telezetts\'{e}g  & obbligo legale & juridisku pienākumu & teisin\.{e} prievol\.{e}\\
  \midrule
  eluline huvi & elint\"{a}rke\"{a} etu & inter\u{e}t vital & leasanna r\'{i}th\'{a}bhachtacha & klju\v{c}ni interes & l\'{e}tfontoss\'{a}g\'{u} \'{e}rdekek & interesse vitale & vitāla interese & gyvybinius interesus\\
  \midrule
  avalik huvi & yleinen etu & int\'{e}r\u{e}t public & leas an phobail & javni interes & k\"{o}z\'{e}rdek & interesse pubblico & sabiedr\~{i}ba interese & vie\v{s}ojo intereso\\
  \midrule
  avalik v\~{o}im & julkinen valta & autorit\'{e} publique & \'{u}dar\'{a}is oifigi\'{u}il & slu\v{z}bene ovlasti & k\"{o}zhatalom & pubblico potere & oficiālās pilnvaras & vie\v{s}osios vald\v{z}ios\\
  \midrule
  avaliku sektori asutus & viranomainen & autorit\'{e} publique & \'{u}dar\'{a}is phoibl\'{i} & javne vlasti  & k\"{o}zhatalmi szervek & autorit\`{a} pubblica & publiskas iestāde & vald\v{z}ios institucija\\

 	\bottomrule
\end{tabular}
\end{table*}

\begin{table*}[ht]
\label{table:wording4}
	\caption{List of GDPR Phrases III}
    \begin{tabular}{p{1.7cm}p{1.7cm}p{1.7cm}p{1.7cm}p{1.7cm}p{1.7cm}p{1.7cm}p{1.8cm}}\\

    \toprule
MT & NL & PL & PT & RO & SK & SL & SV\\

	\toprule

 kontrollur & verwerkings- verantwoordelijke & administrator & respons\'{a}vel pelo tratamento & operator & prev\'{a}dzkovateľ & upravljavec & personuppgiftsansvarige\\
  \midrule
  uffi\.{c}jal tal-protezzjoni tad-data & functionaris voor gegevensbescherming & inspektor ochrony danych & encarregado da prote\c{c}\~{a}o de dados & responsabil protecția datelor; ofițer protecția datelor & zodpovednej osoby & poobla\v{s}\v{c}ena oseba za varstvo podatkov & dataskyddsombud\\
  \midrule
  g\={h}anijiet & verwerkingsdoel & cel & finalidade & scop & \'{u}\v{c}el & namen & syften\\
  \midrule
  ba\.{z}i legali & rechtsgrond  & podstawa prawna & fundamento jur\'{i}dico & temei juridic; baza juridică & pr\'{a}vny z\'{a}klad & pravna podlaga & r\"{a}ttsliga grunden\\
  \midrule
  ippro\.{c}essar & verwerking & przetwarzanie & tratamento & prelucrare & spracovanie & obdelava & behandling\\
  \midrule
  interess leġittimu & gerechtvaardigde belang & uzasadniony interes & interesse leg\'{i}timo & interes legitim & opr\'{a}vnen\'{e} z\'{a}ujmy & zakoniti interes & ber\"{a}ttigade intressen\\
  \midrule
  ri\.{c}evitur & ontvangers & odbiorca & destinat\'{a}rio & destinatar & pr\'{i}jemca & uporabnik & mottagare\\
  \midrule
  pajji\.{z} terz & derde land & państwo trzecie & pa\'{i}s terceiro & țară terță & tretia krajina & tretja dr\v{z}ava & tredjeland \\
  \midrule
  perijodu & periode & okres & prazo de conserva\c{c}\~{a}o & perioada  & doba  & obdobje & period \\
  \midrule
  a\.{c}\.{c}ess & toegang & dost\k{e}p & acesso  & acces & pr\'{i}stup  & dostop  & tillg\r{a}ng \\
  \midrule
  rettifika & rectificatie & sprostowanie & retifica\c{c}\~{a}o  & rectificare & oprava & popravek  & r\"{a}ttelse \\
  \midrule
  t\={h}assir  & wissen & usuni\k{e}cie & apagamento & ștergere & vymazanie  & izbris  & radering \\
  \midrule
  restrizzjoni & beperking & ograniczenie & limita\c{c}\~{a}o  & restricționare & obmedzenie  & omejitev  & begr\"{a}nsning \\
  \midrule
  oġġezzjoni  & bezwaar & wnoszenie sprzeciwu & opor & opune & pr\'{a}vo namietať & ugovarjati & inv\"{a}nda \\
  \midrule
  portabbilt\`{a} tad-data & gegevens- overdraagbaarheid & przenoszenie danych & portabilidade dos dados & portabilitatea datelor & prenosnosť \'{u}dajov & prenosljivost podatkov & dataportabilitet\\
  \midrule
  jiġi irtirat il-kunsens & toestemming intrekken & cofanie zgody & retirar consentimento & retrage consimțământul & s\'{u}hlas odvolať & preklic privolitve & \r{a}terkalla samtycke\\
  \midrule
  ilment & klacht & skarga & reclama\c{c}\~{a}o  & plângere  & sťa\v{z}nosť  & prito\v{z}ba & klagom\r{a}l \\
  \midrule
  awtorit\`{a} supervi\.{z}orja  & toezichthoudende autoriteit & organ nadzorczy & autoridade de controlo  & autoritate de supraveghere & dozorný org\'{a}n & nadzorni organ & tillsynsmyndighet\\
  \midrule
  kuntratt & overeenkomst  & umowa & contrato & contract & zmluva & pogodba & avtal \\
  \midrule
  rekwi\.{z}it statutorju & wettelijke verplichting & wym\'{o}g ustawowy & obriga\c{c}\~{a}o legal  & obligație legală  & z\'{a}konn\'{a} po\v{z}iadavka & statutarna obveznost  & lagstadgat krav\\
  \midrule
  rekwi\.{z}it kuntrattwali & contractuele verplichting & wym\'{o}g umowny & obriga\c{c}\~{a}o contratual & obligație contractuală & zmluvn\'{a} po\v{z}iadavka & pogodbena obveznost  & avtalsenligt krav\\
  \midrule
  konsegwenzi & gevolgen  & konsekwencje  & consequ\u{e}ncias & consecință & n\'{a}sledky  & posledica & f\"{o}ljder\\
  \midrule
  te\={h}id awtomatizzat ta' de\.{c}i\.{z}jonijiet & geautomatiseerde besluitvorming & zautomatyzowane podejmowanie decyzji & decis\~{a}o automatizada & process decizional automatizat & automatizovan\'{e} rozhodovanie & avtomatizirano sprejemanje odlo\v{c}itev & automatiserat beslutsfattande\\
  \midrule
  tfassil tal-profil & profilering & profilowanie & defini\c{c}\~{a}o de perfis & crearea de profiluri & profilovanie & oblikovanje profilov & profilering \\
  \midrule
  jippro\.{c}essa ulterjorment & verdere verwerking & dalsze przetwarzanie & proceder tratamento porterior & prelucrare ulterioară & ďal\v{s}ie spracovanie & nadaljnja obdelava & ytterligare behandla\\
  \midrule
  kunsens & toestemming & zgoda & consentimento & consimțământul & s\'{u}hlas & privolitev & samtycke\\
  \midrule
  twettiq ta' kuntratt & uitvoering van een overeenkomst & wykonanie umowy & execu\c{c}\~{a}o de um contrato & executarea unui contract; execut contract & plnenie zmluvy & izvajanja pogodbe & fullg\"{o}ra ett avtal\\
  \midrule
  obbligu legali & wettelijke verplichting & obowi\k{a}zek prawny & obriga\c{c}\~{a}o jur\'{i}dica & obligație legală & z\'{a}konn\'{a} povinnosť & zakonska obveznost & r\"{a}ttslig f\"{o}rpliktelse\\
  \midrule
  interess vitali & vitale belang & \.{z}ywotny interes & interesse vital & interes vital & \v{z}ivotne dôle\v{z}itý z\'{a}ujem & \v{z}ivljenski interes & vitala intresse\\
  \midrule
  interess pubbliku & algemeen belang & interes publiczny & interesse p\'{u}blico & interes public & verejný z\'{a}ujm & javni interes & allm\"{a}nt intresse\\
  \midrule
  awtorit\`{a} uffi\.{c}jali & openbaar gezag & w\l~adza publiczna & autoridade p\'{u}blica & autoritate publică; autoritatea oficială & verejn\'{a} moc & javna oblast & myndighetsut\"{o}vning\\
  \midrule
  awtorita' pubblika & overheidsinstantie & organ publiczny & autoridades p\'{u}blicas & autoritate publică & org\'{a}n verejnej moci & javni organ & offentlig myndighet\\

 	\bottomrule
\end{tabular}
\end{table*}


\end{document}